\documentclass[twocolumn]{aastex631}
\usepackage[utf8]{inputenc}
\usepackage{amsmath, amsthm, amssymb}
\usepackage{graphics}
\usepackage{subfigure}
\usepackage{indentfirst}
\usepackage{enumerate}
\usepackage{url}
\usepackage{hyperref}
\usepackage{booktabs}
\usepackage[figuresright]{rotating}
\usepackage{ragged2e}
\usepackage{soul}

\newcommand{\ha}{\hbox{H$\alpha$}}
\newcommand{\hii}{\hbox{H\,{\sc II}}}
\newcommand{\nii}{\hbox{[N\,{\sc II}]}}
\newcommand{\sii}{\hbox{[S\,{\sc II}]}}
\newcommand{\oii}{\hbox{[O\,{\sc II}]}}
\newcommand{\oiii}{\hbox{[O\,{\sc III}]}}

\begin{document}

    \title{Galactic \hii\ regions in LAMOST Medium-Resolution Spectroscopic Survey of Nebulae}
    
    \author[0009-0000-6954-9825]{Yunning Zhao}
    \affiliation{National Astronomical Observatories,  Chinese Academy of Sciences, Beijing 100101, China}
    \affiliation{School of Astronomy and Space Science, University of Chinese Academy of Sciences, Beijing 100049, China}

    \author[0000-0002-1783-957X]{Wei Zhang{$^\dagger$}}
    \affiliation{National Astronomical Observatories,  Chinese Academy of Sciences, Beijing 100101, China}
    \email{Wei Zhang{$^\dagger$}, xtwfn@bao.ac.cn}
    
    \author{Lin Ma}
    \affiliation{{The Key Laboratory of Cosmic Rays (Tibet University), Ministry of Education, Lhasa 850000, China}}
    \affiliation{National Astronomical Observatories,  Chinese Academy of Sciences, Beijing 100101, China}
    \affiliation{School of Astronomy and Space Science, University of Chinese Academy of Sciences, Beijing 100049, China}
    
    \author[0009-0008-1361-4825]{Shiming Wen}
    \affiliation{National Astronomical Observatories,  Chinese Academy of Sciences, Beijing 100101, China}
    
    \author{Tao Jing}
    \affiliation{Department of Astronomy, Tsinghua University, Beijing 100084, China}
    
    \author{Cheng Li}
    \affiliation{Department of Astronomy, Tsinghua University, Beijing 100084, China}
    
    \author{Zheng Zheng}
    \affiliation{State Key Laboratory of Radio Astronomy and Technology, National Astronomical Observatories, Chinese Academy of Sciences, Beijing 100101, China}
    
    \author{Aiyuan Yang}
    \affiliation{National Astronomical Observatories,  Chinese Academy of Sciences, Beijing 100101, China}
    \affiliation{State Key Laboratory of Radio Astronomy and Technology, National Astronomical Observatories, Chinese Academy of Sciences, Beijing 100101, China}
    
    \author[0009-0004-1948-2158]{Shichao Han}
    \affiliation{National Astronomical Observatories,  Chinese Academy of Sciences, Beijing 100101, China}

    \author{Juanjuan Ren}
    \affiliation{National Astronomical Observatories,  Chinese Academy of Sciences, Beijing 100101, China}

    \author{Jianjun Chen}
    \affiliation{National Astronomical Observatories,  Chinese Academy of Sciences, Beijing 100101, China}

    \author{Hong Wu}
    \affiliation{National Astronomical Observatories,  Chinese Academy of Sciences, Beijing 100101, China}

    \author{Yongheng Zhao}
    \affiliation{National Astronomical Observatories,  Chinese Academy of Sciences, Beijing 100101, China}

\begin{abstract}

 Based on LAMOST Medium-Resolution Spectroscopic Survey of Nebulae (MRS-N) data and WISE Galactic \hii\ region catalog, we construct a sample of 280 Galactic \hii\ regions and candidates in the Outer Galaxy (80$^{\circ}$ $\lesssim$ l $\lesssim$ 220$^{\circ}$). Using MRS-N optical spectra, we measure four emission lines (\ha, \nii$\lambda$6584, \sii$\lambda\lambda$6717,6731) and use line-ratios to spectroscopically confirm 255 \hii\ regions, including 90 previously ``Known" \hii\ regions and 165 newly classified ones.
 We measure their $T_{\rm e}$, $n_{\rm e}$ and oxygen abundance, and determine distances via associated OB stars and the kinematic method. The sample spans $R_{\rm gal}$ from 8.16 to 15.36 kpc, enabling investigation of radial gradients in physical properties. 
 We find \nii/\ha\ and \sii/\ha\ decrease with increasing $R_{\rm gal}$, while \sii/\nii\ remains nearly flat; these trends are quite different from diffuse ionized gas (DIG). We derive the $T_{\rm e}$ gradient of 344.530 $\pm$ 78.083 K kpc$^{-1}$, and the $\log n_{\rm e}$ gradient of -0.143 $\pm$ 0.041 cm$^{-3}$ kpc$^{-1}$. Oxygen abundance shows a steep slope of -0.044 $\pm$ 0.010 dex kpc$^{-1}$ in the inner disk and a shallow slope of -0.016 $\pm$ 0.005 dex kpc$^{-1}$ in the outer disk, with a global slope of -0.014 $\pm$ 0.005 dex kpc$^{-1}$.
 We also examine the two-dimensional distributions of $T_{\rm e}$, $n_{\rm e}$, and oxygen abundance, and find the gradients vary with azimuth. There is no obvious difference between spiral arm and interarm regions, and no trend appears along individual arms.
 From \nii/\ha\-\sii$\lambda$6717/\ha\ diagram, \hii\ regions have a S$^+$/S ratio (0.32), lower than DIG (0.43); however, heavy overlap prevents clear separation from this diagram alone.

\end{abstract}


\section{Introduction} \label{sec:intro}
 H II regions are ionized nebulae formed when ultraviolet (UV) radiation from young massive stars photoionizes the surrounding interstellar medium (ISM). 
 They emit strongly across optical, infrared (IR), and radio wavelengths, making them detectable throughout the Galactic disk, although their optical brightness can be significantly attenuated by dust extinction. Their spectra are characterized by strong emission lines of hydrogen and various metals \citep{Baldwin1981PASP}, which are used to derive key astrophysical properties such as star formation rates, internal reddening, chemical abundances, electron temperature, and electron densities \citep{Perez-Montero2017PASP}.
 Because the ionization sources are short-lived stars and most of the gas is swept away by radiation pressure, the typical age of \hii\, regions is only $\sim$10 Myr \citep{Kennicutt1988AJ}, making them the tracers of star formation at the present epoch.
 \hii\, regions have been extensively studied in a variety of contexts, including the Galactic spiral structure \citep{Georgelin1976AA, Caswell1987AA, Russeil2003AA, Hou2009AA, Reid2019ApJ, Shen2025A&A},
 the abundance distribution across the disk and chemical evolution of the Milky Way \citep{Shaver1983MNRAS, Deharveng2000MNRAS, Quireza2006ApJ, Rudolph2006ApJS, Balser2011ApJ, Wenger2019ApJ, Arellano_Cordova2020MNRAS, Mendez_Delgado2022MNRAS},
 the scaling relations in kinematics \citep{Larson1981MNRAS, Wisnioski2012MNRAS, Cosens2018ApJ, Cosens2022ApJ, Ma2026RAA},
 and stellar feedback during their expansion \citep{Lopez2011ApJ, Lopez2014ApJ, Barnes2020MNRAS, Pathak2025ApJ}.

 Owing to its significance, consistent efforts have been made across multiple wavelengths to build large \hii\ region samples.
 In the mid-infrared (MIR), \cite{Anderson2014ApJS} constructed a Galactic \hii\, region catalog based on data from the Wide-Field Infrared Survey Explorer (WISE), using the MIR morphology of \hii\, regions. This catalog contains 8400 sources, of which only 1524 are identified as ``Known" \hii\ regions--those with detected \ha\, emission or radio recombination lines (RRLs)--while the majority of the remaining candidates still await confirmation. 
 In the radio, extensive surveys of radio continuum and RRLs have further expanded the number of Galactic \hii\, regions, including the Southern \hii\, Region Discovery Survey \citep[SHRDS,][]{Wenger2021ApJS}, the GLOSTAR Galactic plane survey \citep[e.g.,][]{Yang2023A&A680A92Y, Khan2024A&A}, and the SARAO MeerKAT survey \citep[e.g.,][]{Bordiu2025A&A, Mutale2026MNRAS546f1849M}, as well as the survey of the densest and most compact \hii\ regions, i.e., hypercompact \hii\ regions \citep[e.g.,][]{Yang2019MNRAS, Yang2021AA645A110Y, Patel2023MNRAS, Patel2024MNRAS, Patel2025MNRAS, Yang2025AA694A26Y}.
 In the optical, \cite{Madsen2006ApJ} reported spectral observations of 15 large Galactic \hii\ regions using the Wisconsin \ha\, Mapper (WHAM), which has surveyed the distribution and kinematics of ionized gas in the northern sky \citep{Haffner2003ApJS}.
 However, WHAM has a low spatial resolution ($\sim$ 1\arcdeg), limiting its ability to resolve \hii\, regions with small angular sizes.
 Additional optical studies have been conducted using large-aperture telescopes \citep{Esteban2013MNRAS, Fernandez_Martin2017A&A, Arellano_Cordova2021MNRAS}, all of which involve sample sizes around dozens. These observations are typically time-intensive due to the high cost of large-telescope time, resulting in limited sample sizes, and more generally, constrained by the long-slit observation mode, the numbers of \hii\ regions observed have not been significantly expanded. 
 
 LAMOST, also known as the Guo Shou Jing Telescope, is a 4-meter quasi-meridian reflective Schmidt telescope located at Xinglong Observatory in China. It is equipped with 4000 fibers on its focal plane and  has a field of view (FOV) of 5$^{\circ}$. The advantages of large FOV and multiple targets allow LAMOST to be used for large-scale spectroscopic surveys \citep{Wang1996ApOpt, Su2004ChJAA, Cui2012RAA, Zhao2012RAA, Luo2015RAA}. \cite{Wang2018PASP} spectroscopically identified 101 Galactic \hii\ regions using low-resolution spectra of LAMOST. \cite{Yang2021PASP} and \cite{Wang2023ApJS} developed novel machine-learning methods and searched 89 and 57 \hii\ regions, respectively. \cite{Lu2022RAA} used the Product Quantization (PQ) based approximate nearest neighbor (ANN) search to obtain the nearest neighbors of each spectrum, and identified 29 new \hii\ regions. These studies have demonstrated LAMOST's capability in detecting \hii\ regions and its high efficiency in constructing large catalogs.
 
 LAMOST Medium-Resolution Spectroscopic Survey of Galactic Nebulae (MRS-N) is a specific survey focused on Galactic ionized nebulae. Based on this dataset, \cite{Wen2025AJ} compiled a large sample of diffuse ionized gas (DIG) in the anti-center region of the Milky Way and analyzed the radial and vertical distributions of three line ratios (\nii/\ha, \sii/\ha, and \sii/\nii) along with oxygen abundance. \cite{Zhang2025AJ} selected a sample of 10 isolated \hii\ regions and examined the 1D radial profiles of emission line fluxes (\ha, \nii, and \sii), line ratios (\nii/\ha, \sii/\ha, and \sii/\nii), and radial velocities of these three emission lines. They subsequently investigated the evolution of the escape fraction during \hii\ region expansion. Using this small \hii\ region sample, \cite{Ma2026RAA} found that these small-size \hii\ regions remarkably follow the same size–velocity dispersion relation as giant \hii\ regions. In this work, we plan to construct a large sample of \hii\ regions from the LAMOST MRS-N dataset and perform a statistical analysis of their properties.
 
 This paper is organized as follows: Section 2 describes the LAMOST MRS-N observations and sample selection criteria; Section 3 presents the procedures of spectra reduction and emission-line measurements; Section 4 reports our main results including electron temperature ($T_{\rm e}$), electron density ($n_{\rm e}$), oxygen abundance, and distances of the \hii\ regions; Section 5 analyzes the radial gradients and two-dimensional distributions of line ratios, $T_{\rm e}$, $n_{\rm e}$ and oxygen abundance, and discusses the distinction between \hii\ regions and DIG. Finally, Section 6 summarizes the main findings and conclusions of this work.

\section{Data and sample selection} \label{sec:data and sample selection}

\subsection{LAMOST MRS-N} \label{sec:MRS-N}

 LAMOST has been conducting its Medium-Resolution Spectroscopic Survey (MRS) since September 2018. It is equipped with 16 medium spectral resolution (R $\sim$ 7500) spectrographs, each covering a blue band (4950 \AA\ $\sim$ 5350 \AA) and a red band (6300 \AA\ $\sim$ 6800 \AA) \citep{Liu2020arXiv}. LAMOST MRS-N is a sub-section of LAMOST MRS. This survey aims to map Galactic nebulae including \hii\ regions, supernova remnants (SNRs), planetary nebulae (PNe), DIG, and some bright stars. LAMOST MRS-N plans to cover approximately 1700 square degrees, spanning the northern Galactic plane in $40^{\circ} < l < 215^{\circ}$ and $|b| < 5^{\circ}$ \citep{Wu2021RAA}. In the MRS-N survey, the observing strategy divided the field of view into $2\arcmin \times 2\arcmin$ grids, and fibers were assigned either to bright stars with $r < 13$ mag or, otherwise, to the brightest H$\alpha$-emitting positions detected in IPHAS images, resulting in a target-driven and spatially non-uniform fiber distribution. The spectra cover some emission lines of \hii\ regions, such as \ha, \nii, \sii, and \oiii, which are crucial to understanding the physical properties of emission nebulae. For this survey, the wavelengths have been recalibrated by fitting seven sky emission lines, and the precision of radial velocity (RV) is better than 1 $\rm km\,s^{-1}$ \citep{Ren2021RAA}. The geocoronal \ha\, emission lines (H$_{\alpha,\rm sky}$) have been subtracted by a new method, which uses the linear correlation between the line ratio of H$_{\alpha,\rm sky}$ to the skyline OH$\lambda$6554 and solar altitude \citep{Zhang2021RAA}. The whole data processing pipeline and data products are summarized in \cite{Wu2022RAA}.

\begin{figure*}[htbp]
    \centering
    \includegraphics[width=\linewidth]{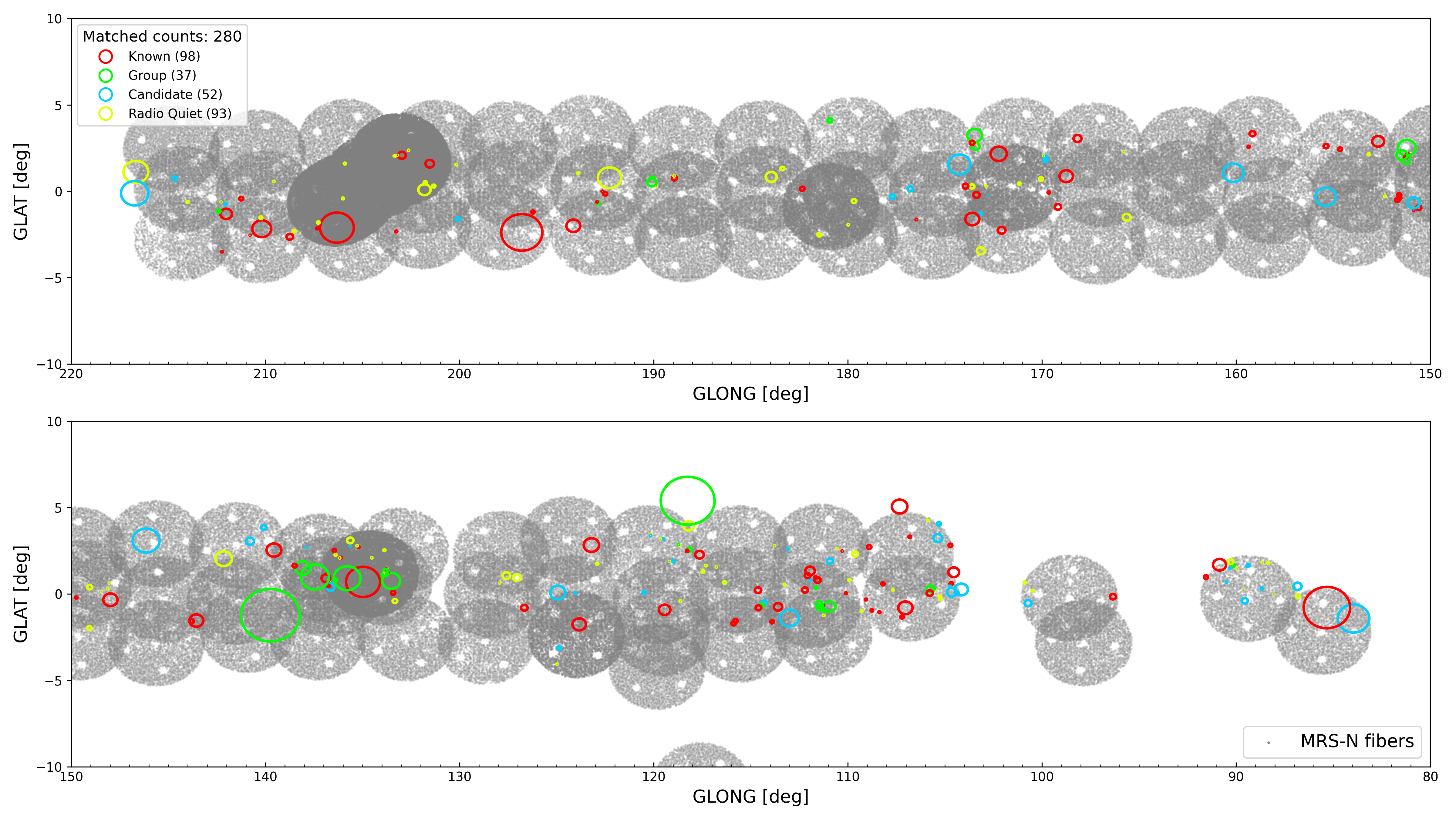}    
    \caption{Spatial distribution of MRS-N observations and our sample constructed in Section \ref{sec:sample}. Gray dots denote all MRS-N fibers without S/N cuts. Circles of different colors represent sources of different catalogs: red, green, blue, and yellow indicate Known (`K'), Group (`G'), Candidate (`C'), and radio-Quiet (`Q'), respectively.}
    \label{fig:coverage}
    \end{figure*}

 By February 2024, LAMOST MRS-N had completed 119 plate observations, which yielded a dataset of 480,000 spectra. Figure \ref{fig:coverage} shows the footprint of this survey in Galactic longitude and latitude, where each gray dot represents a single fiber. This footprint includes all fibers and has not been filtered by signal-to-noise (S/N).

\subsection{WISE HII region catalog}

 \cite{Anderson2014ApJS} constructed a catalog of Galactic \hii\ regions and candidates based on the MIR morphology characteristics observed by the all-sky WISE satellite. The catalog is constantly updated on its online website (\url{http://astro.phys.wvu.edu/wise/}). 

 In this work, we use the latest version (V2.3) of the WISE \hii\ catalog (hereafter HIICat\_V2.3), released in December 2020. HIICat\_V2.3 contains 8416 sources and provides WISE\_Name ID, coordinates, catalogs, and angular radius for each source.
 There are five source catalogs defined by HIICat\_V2.3: ``Known" sources have measured RRL or \ha\ emission; ``Group" sources are positionally associated with known \hii\ region complexes; ``Candidate" sources are spatially coincident with radio continuum emission, but lack RRL or \ha\ observation; ``radio-Quiet" sources show no radio continuum emission; and ``?" sources lack high-quality radio continuum data. 
 
 Additionally, Using data from the literature, this catalog has collected distances for 1513 sources, with values derived from two methods: trigonometric parallaxes of associated masers and kinematic distances.

\subsection{Sample selection} \label{sec:sample}

    \begin{figure*}[htbp]
    \centering
    \includegraphics[width=\linewidth]{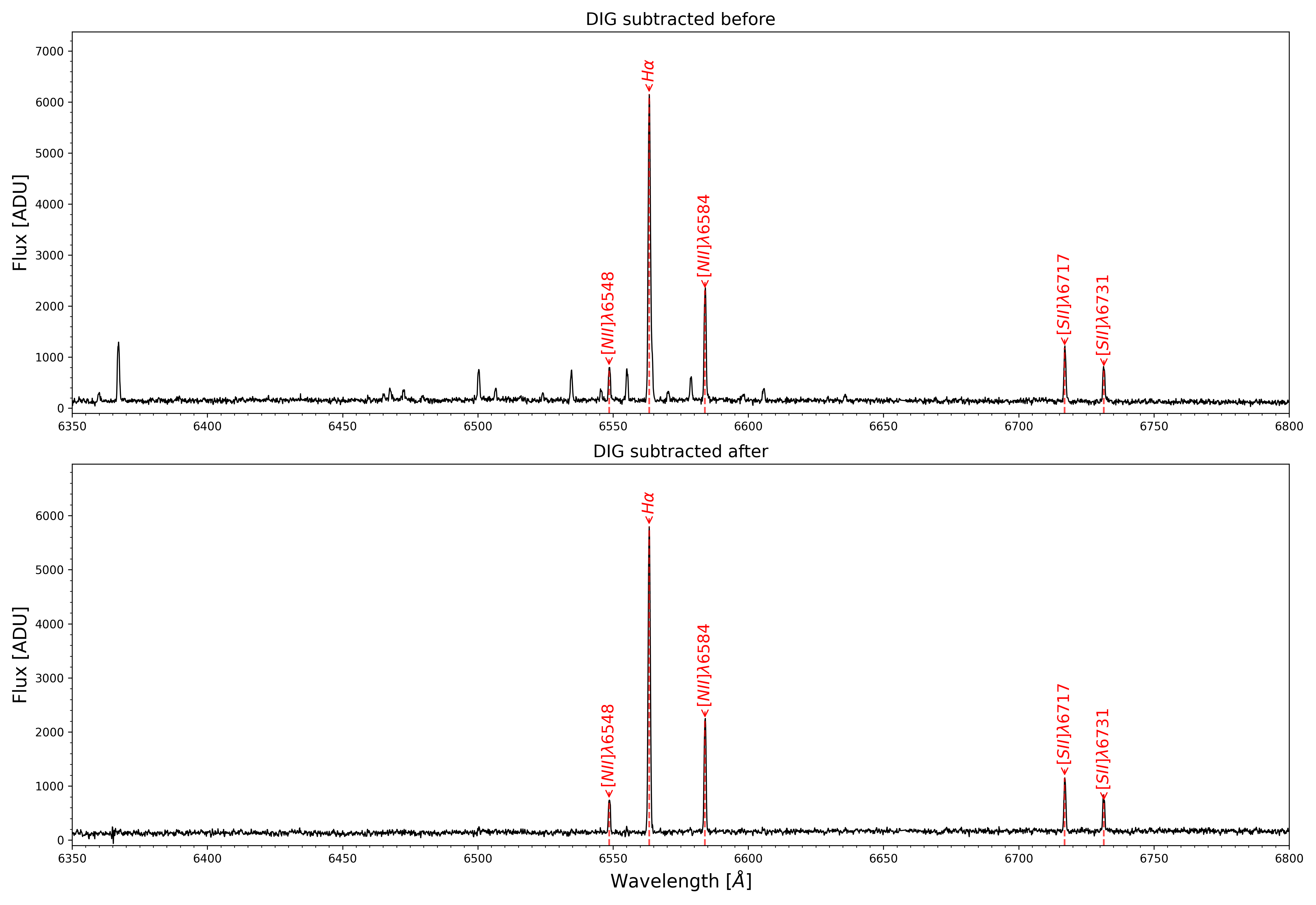}    
    \caption{Example of the sky/DIG subtraction procedure. The observed spectrum before sky/DIG subtraction is shown in the upper panel, while the sky/DIG-subtracted spectrum is shown in the lower panel. The main emission lines are marked with red dashed vertical lines. The flux is given in units of ADU.}
    \label{fig:examplespec}
    \end{figure*}

 We cross-match the MRS-N dataset with HIICat\_V2.3. and select \hii\ regions and candidates that have at least one spectrum in the angular radius of the source. Additionally, if one \hii\ region is entirely located within the radius of another, only the outer \hii\ region is retained. Following these steps, we obtain a parent sample containing 309 \hii\ regions and candidates.
 
 Since \hii\, region spectra are often contaminated by foreground/background emission from DIG, we reduce the spectra following the method described in \cite{Ma2026RAA} to remove the DIG component from the observed spectra. Specifically, we select fibers located outside \hii\ regions with the lowest \nii$\lambda$6584 emission flux to construct representative DIG spectra, and then subtract these from the observed spectra.  We show example spectra before and after sky/DIG subtraction in Figure \ref{fig:examplespec}. The sky/DIG-subtracted spectrum after subtraction has all the skylines and DIG emission removed. We only retain those spectra where the S/N for the four main emission lines (\ha, \nii$\lambda$6584, \sii$\lambda\lambda$6717,6731) are greater than 3.

 The final sample consists of 280 \hii\ regions and candidates. Figure \ref{fig:coverage} shows their distribution, with circles color-coded by HIICat\_V2.3 catalogs: red (Known, K), green (Group, G), blue (Candidate, C), and yellow (radio-Quiet, Q). 
 Among our sample, 98 are classified as ``K", 37 as ``G", 52 as ``C", and the remaining 93 as ``Q".  In the next section, we will further identify \hii\ regions from this sample using LAMOST MRS-N optical spectra.

\section{Spectra reduction, measurements, and classification} \label{sec:spectra}

\subsection{Relative flux calibration}

 As the current MRS-N spectra have not been performed absolute flux calibration, we carry out relative flux calibration following the procedure in \cite{Ma2026RAA}. This method uses the skyline OH$\lambda$6554 to do the flux alignment for each plate. In a plate containing $N$ fibers, the corrected flux for the $i$-th fiber is as follows:
    \begin{equation}
        F_{i}(neb) = 
        \frac{F_{i}^{o}(neb)}{F_{i}^{o}(6554)} 
        \frac{\sum F_{i}^{o}(6554)}{N}
    \end{equation}
where $F_{i}^{o}(neb)$ represents the observed flux of nebular emission lines, including \ha, \nii\,$\lambda$6584, and \sii\,$\lambda\lambda$6717,6731. $F_{i}^{o}(6554)$ represents the observed flux of the skyline OH$\lambda$6554.

\subsection{Spectra stacking and measurement}\label{sec:spec process}

    \begin{figure*}[htbp]
    \centering
    \includegraphics[width=0.95\linewidth]{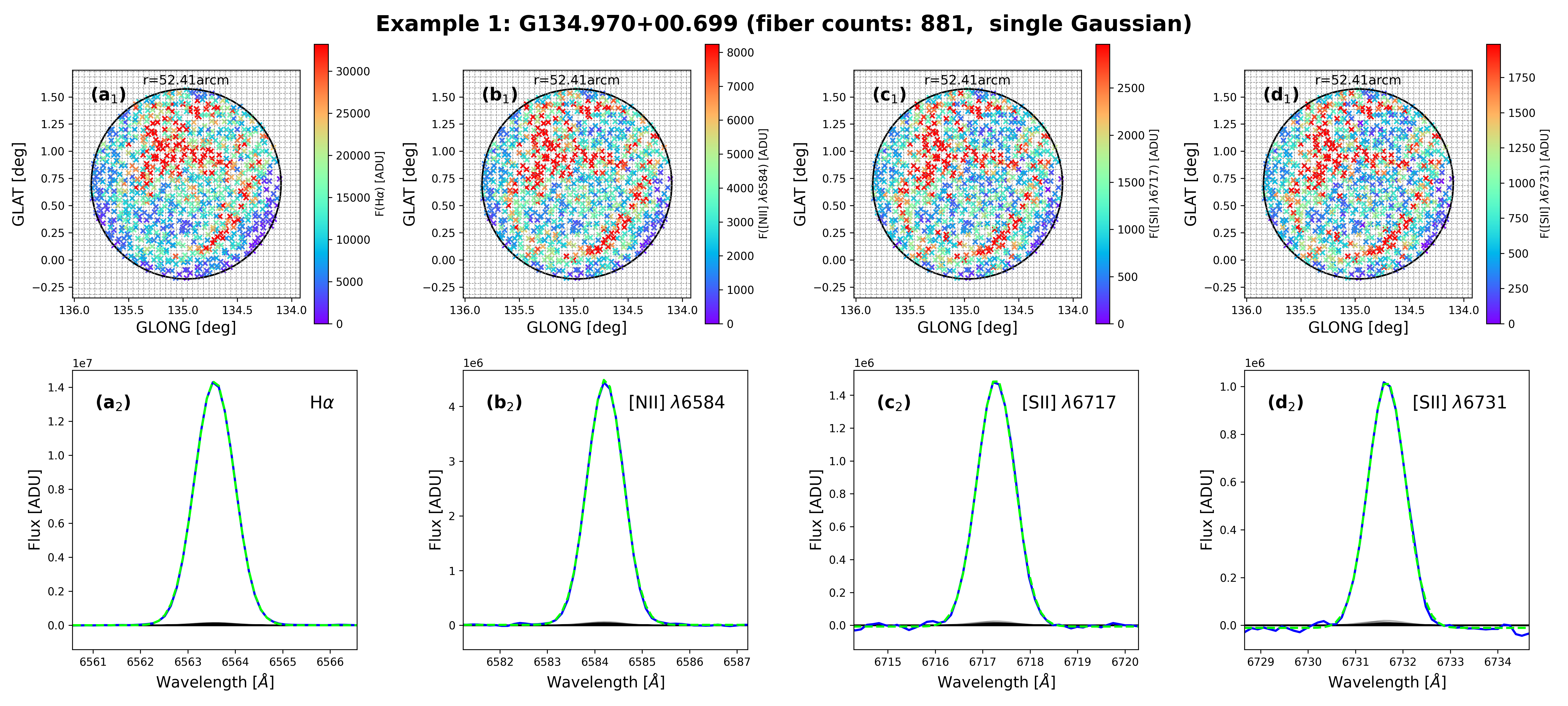}    
    \vspace{\baselineskip}
    \includegraphics[width=0.95\linewidth]{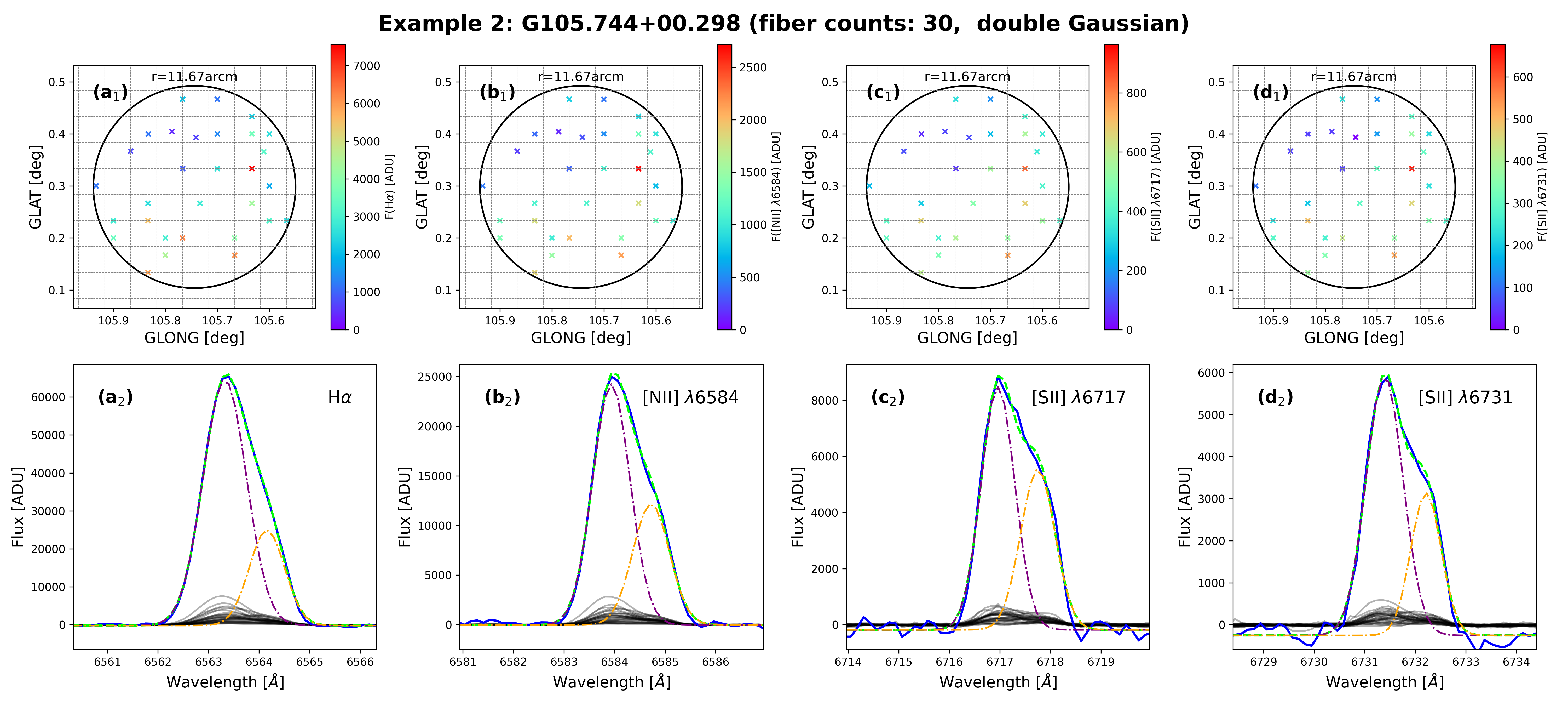}
    \caption{Fiber distributions and spectral processes for two example \hii\ regions.
    \textbf{\textit{Example 1}:} 
    Panels ($a_1$) $\sim$ ($d_1$) show the spatial distributions of fibers in the radius range, overlaid with 3$\arcmin$ $\times$ 3$\arcmin$ grids. The colorbars indicate the flux values of \ha, \nii$\lambda6584$, \sii$\lambda6717$, and \sii$\lambda6731$, respectively. Panels ($a_2$) $\sim$ ($d_2$) show the processes of stacking and Gaussian fitting. Each gray line represents one individual spectrum, and the blue lines are the stacked spectra. The green dashed lines are the Gaussian fitting results.
    \textbf{\textit{Example 2}:} 
    Same as Example 1, and the additional purple and orange dashed lines represent the two Gaussian components.}
    \label{fig:gaussfit}
    \end{figure*}

 In the MRS-N survey, some fields are observed multiple times and there are overlaps between nearby plates, causing some \hii\ regions and candidates to be oversampled. Therefore, resampling the data to achieve a uniform density is necessary. The spatial resolution (defined by the median fiber-to-fiber distance) of MRS-N is approximately 3$\arcmin$ \citep{Wen2025AJ}. So we resample the data into 3$\arcmin$ $\times$ 3$\arcmin$ bins. When multiple spectra fall within a single bin, we retain only the one with the highest S/N of the \ha\, emission line.

 For each source in our sample, we stack the resampled spectra within its radius and apply Gaussian fitting to the stacked spectra.  For each stacked spectrum, we first use a single-Gaussian profile. For a fraction of the spectra (about 19\%), the fitting residual is large, so a double-Gaussian profile is employed to yield an acceptable residual.  Figure \ref{fig:gaussfit} presents two examples for these two cases. 

 With this method, we obtain the emission line information including integrated flux, full width at half maximum (FWHM), and RV. 
 Since absolute flux calibration has not been performed, we only focus on the line ratios rather than fluxes.
 In this paper, we define F(\ha), F(\nii), and F(\sii) as the integrated fluxes of \ha, \nii$\lambda$6584, and \sii$\lambda\lambda$6717,6731 emission lines. \nii/\ha, \sii/\ha, and \sii/\nii\ are defined as F(\nii)/F(\ha), F(\sii)/F(\ha), and F(\sii)/F(\nii), respectively. Additionally, W$_{\rm H\alpha}$ and W$_{\rm [N II]}$ represent the FWHMs of \ha\, and \nii$\lambda$6584 in units of $\rm km\,s^{-1}$. 
 The W$_{\rm H\alpha}$ and W$_{\rm [N II]}$ have been corrected for instrumental broadening. We perform single-Gaussian fits on OH $\lambda$ 6554 skylines within the 1R range of each \hii\ region and adopted the median fitted width as the instrumental broadening. 
 The local standard of rest (LSR) velocity ($V_{\rm LSR}$) is obtained by converting RV of \ha\, using the motion of the Sun with respect to the LSR. The Sun moves at a speed of about 20 $\rm km\,s^{-1}$ toward (RA = 18h03m50.29s, DEC = +30\arcdeg00\arcmin16.8\arcsec) (J2000).

\subsection{Identification of \hii\ Regions} \label{sec:identification}

 Besides \hii\ regions, PNe and SNRs usually exhibit similar emission lines, but their line ratios are significantly different. 
 There are several empirical emission-line diagnostic diagrams to classify ionized nebulae in the Galaxy. The diagnostic diagram of Sabbadin-Minello-Bianchini (SMB), originally proposed by \cite{Sabbadin1977A&A}, demonstrates how to separate \hii\ regions from PNe and SNRs using \nii/\ha\, and \sii/\ha\, ratios. The SMB diagram has been commonly used to distinguish among different Galactic ionized nebulae \citep{Riesgo_Tirado2002RMxAC, Magrini2003A&A, Kniazev2008MNRAS, Lagrois2012MNRAS, Parker2022FrASS}. In particular, SMB diagrams have been proven useful for identifying \hii\ regions from LAMOST low-resolution spectra in recent years \citep{Wang2018PASP, Zhang2020RAA, Lu2022RAA}. 
 
\begin{figure}[htbp]
    \centering
    \includegraphics[width=\linewidth]{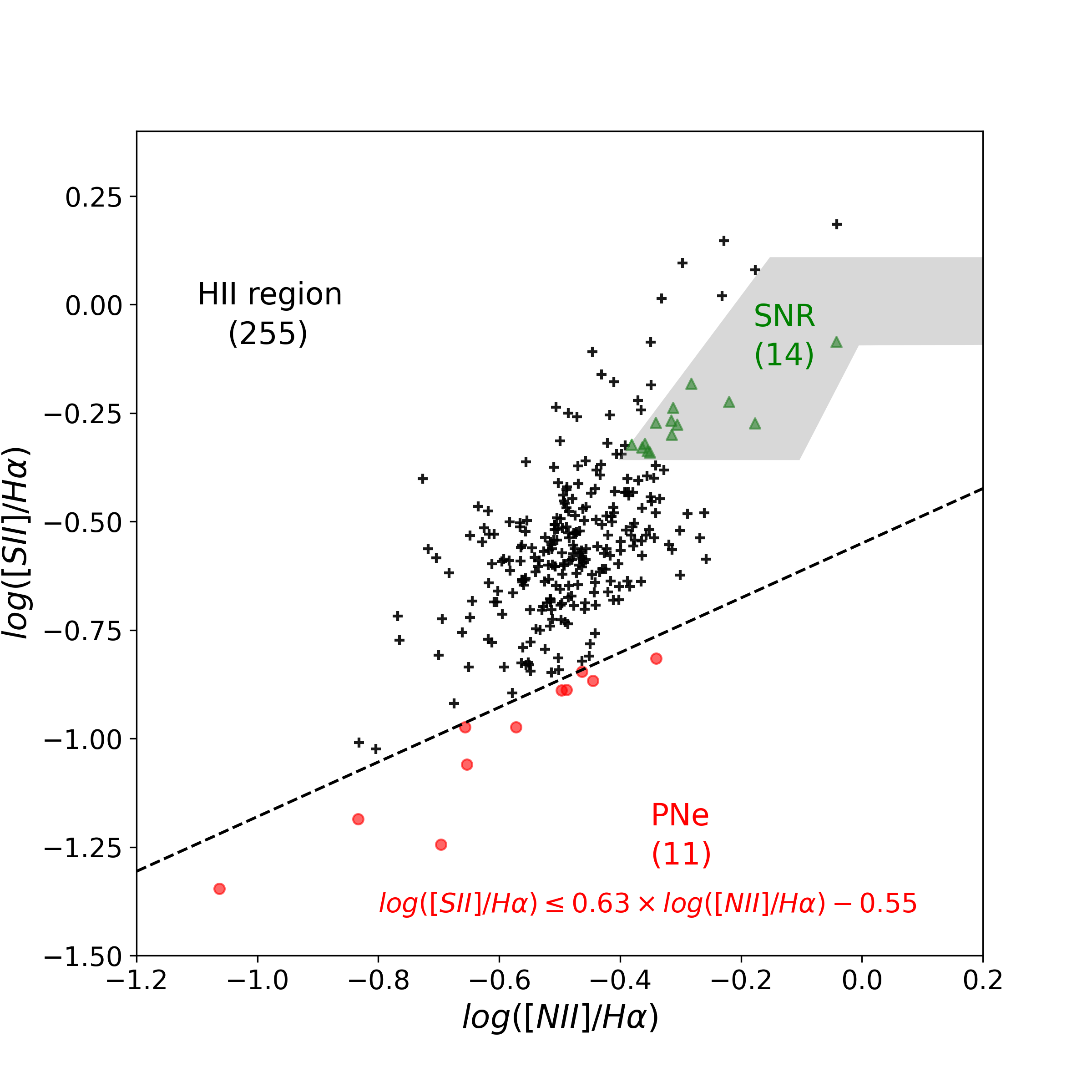}
    \caption{SMB diagnostic diagram. 
    The black dashed line is the criterion to separate \hii\ regions and PNe, whose mathematical expression is shown in the lower right corner. The gray shaded area belongs to SNRs. \hii\ regions, PNe, and SNRs are denoted by black crosses, red dots, and green triangles, respectively.}
    \label{fig:diagnostic} 
    \end{figure}

 Here, we employ the classification criteria proposed by \cite{Kniazev2008MNRAS} for our sample. The diagnostic diagram is shown in Figure \ref{fig:diagnostic}.
 Among these sources, 255 sources fall within the boundaries of \hii\ regions (black crosses), of which 90 are previously ``Known" \hii\ regions and 165 \hii\ region candidates are classified as \hii\ regions. 11 sources are classified as PNe candidates (red dots), and 14 sources are classified as SNRs candidates (green triangles). In the subsequent analysis, we only focus on \hii\ regions classified in this section. 
 
\begin{figure*}[htbp]
    \centering
    \includegraphics[width=0.9\linewidth]{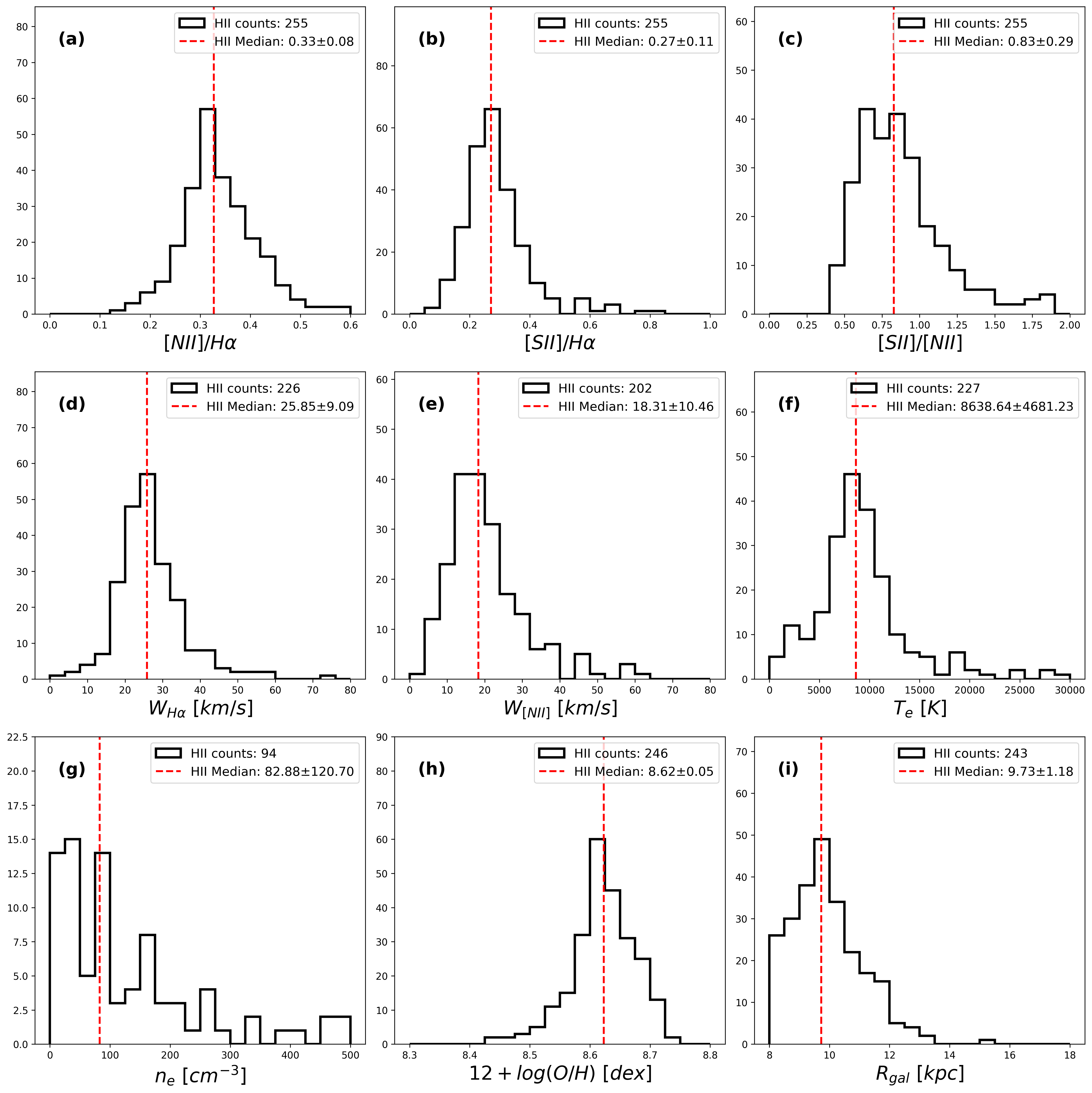}    
    \caption{
    Histograms of parameters. The valid counts are shown in the upper right corner. The red dashed vertical lines indicate the median values, with their corresponding uncertainties also provided.
    }
    \label{fig:parahist} 
    \end{figure*}

 In Figure \ref{fig:parahist}, panels (a) to (e), we present the histograms of \nii/\ha, \sii/\ha, \sii/\nii, $W_{\rm H\alpha}$, and $W_{[\rm N II]}$ for the \hii\ regions. Detailed values are provided in Table \ref{tab:basic info}.

\section{Physical properties} \label{sec:results}

\subsection{Electron temperature}\label{sec:Te}

 The electron temperature ($T_{\rm e}$) of an \hii\ region in thermal equilibrium results from the balance between heating and cooling mechanisms. 
 Multiple physical factors can influence $T_{\rm e}$, including the effective temperature of ionizing source, electron density, dust grains, and heavy-element abundance \citep{Quireza2006ApJ}. Several approaches exist to estimate $T_{\rm e}$ from optical emission lines, such as using line widths, auroral line ratios, and variations in \nii/\ha\, and \oii/\ha\, ratios \citep{Haffner2009RvMP}.
 
 Benefiting from the spectral resolution of MRS-N (R $\sim$ 7500), we can derive $T_{\rm e}$ from the difference between the line widths of hydrogen and nitrogen lines, following \cite{Reynolds1977ApJ}:  
    \begin{equation}
        T_{\rm e} {\rm [K]}= 23.5W_{\rm H\alpha,obs}^2 \times (1- \frac{W_{\rm [N II], obs}^2}{W_{\rm H\alpha,obs}^2}) 
    \end{equation}
 where $W_{\rm H\alpha,obs}$ and $W_{\rm [N II],obs}$ denote the observed FWHMs (in $\rm km\,s^{-1}$) of \ha\, and \nii$\lambda$6584. Here, we use the observed FWHMs without subtracting instrumental broadening, because \ha\, and \nii\, are close in wavelength and their instrumental broadenings are similar, so the instrumental terms approximately cancel in the equation.
 
 We derive $T_{\rm e}$ values for 227 of the 255 \hii\, regions. The remaining 28 \hii\ regions cannot yield reliable $T_{\rm e}$ by this method because their $W_{\rm H\alpha}$ values are less than $W_{\rm [N II]}$.  As shown in panel (f) of Figure \ref{fig:parahist}, the $T_{\rm e}$ values are concentrated around 8000 K, consistent with those of typical \hii\, regions \citep{Lequeux2005ism}. Detailed values are listed in Table \ref{tab:parameters}.

\subsection{Electron density}\label{sec:ne}

 Electron density ($n_{\rm e}$) is another essential parameter of an \hii\, region. We adopt the \sii$\lambda$6717/$\lambda$6731 density-sensitive line ratios to measure $n_{\rm e}$, following \cite{Proxauf2014A&A}:
    \begin{equation}
        R = \frac{F([\rm S\,II] \lambda 6717)}{F(\rm [S\,II] \lambda 6731)}
    \end{equation}
    \begin{equation}
    \begin{aligned}
        {\rm log}(n_{\rm e}\,[{\rm cm^{-3}}]) =  0.0543\times{\rm tan}(-3.0553R+2.8506)  \\
              + 6.98 -10.6905R + 9.9186R^2 -3.5442R^3
    \end{aligned}
    \end{equation}
 The equation is valid for $R < 1.42$.

 We estimate $n_{\rm e}$ values for 94 of the 255 \hii\ regions. The remaining 161 \hii\ regions are beyond the applicable range of the equation. The histogram is shown in panel (g) of Figure \ref{fig:parahist}, with detailed values listed in Table \ref{tab:parameters}.

\subsection{Oxygen abundance}

 There are two main methods to determine oxygen abundance: direct method (also known as $T_{\rm e}$ method) and strong-line method. 
 The direct method requires precise measurements of $T_{\rm e}$ based on the auroral emission lines (e.g., \oiii$\lambda$4363 and \nii$\lambda$5755) \citep{Shi_2005, Kewley_2019}. Although this method is reliable for metallicity estimation, the auroral lines are generally too weak to be detected, especially in high‑metallicity regions. To address this problem, many alternative methods using strong lines have been developed, such as $R_{\rm 23}$ \citep{Pagel1979MNRAS, Kobulnicky2004ApJ}, $N_{\rm 2}\ha$ \citep{van_Zee1998AJ, Pettini2004MNRAS, Marino2013A&A}, $N_{\rm 2}S_{\rm 2}\ha$ \citep{Dopita2016ApSS}, $N_{\rm 2}O_{\rm 2}$ \citep{Kewley2002ApJS}, and $O_{\rm 3}N_{\rm 2}$ \citep{Alloin1979A&A, Pettini2004MNRAS, Marino2013A&A}.  These strong-line methods are calibrated empirically either from direct metallicities or theoretical photoionization models. 
 
 We calculate the oxygen abundance using the $N_{\rm 2}\ha$ diagnostic. We do not employ the direct method because the required auroral lines fall outside the wavelength coverage of our spectra. Other strong-line methods such as $R_{\rm 23}$, $N_{\rm 2}O_{\rm 2}$ and $O_{\rm 3}N_{\rm 2}$, all involve \oii$\lambda\lambda$3727, 3729, \oiii$\lambda\lambda$4959, 5007 or H$\beta$, which are also outside our wavelength range.
 We use the $N_{\rm 2}\ha$ diagnostic rather than the $N_{\rm 2}S_{\rm 2}\ha$ diagnostic because the $N_{\rm 2}\ha$ diagnostic utilizes only \ha\, and \nii$\lambda$6584, which are close in wavelength, thereby minimizing the effects of  extinction correction. Another reason is to enable comparison with results from other studies that also employ this method. The simplicity of relying on only two strong lines makes it one of the most popular diagnostics.
 We use the expression calibrated by \cite{Pettini2004MNRAS}:
    \begin{equation}
    \rm
        12 + log(O/H) = 8.90 + 0.57 \times \log (\nii/\ha)
    \end{equation}
 This equation is usually valid for $\rm -2.5 < log(\nii/\ha) < -0.3$.
 
 We derive oxygen abundances for 246 of the 255 \hii\ regions in our sample, with values ranging from 8.43 to 8.73 dex. Oxygen abundances are not calculated for the remaining 9 \hii\ regions, as their \nii/\ha\ values are beyond the valid range. The histogram is shown in panel (h) of Figure \ref{fig:parahist}, and detailed values are listed in Table \ref{tab:parameters}.


\begin{table*}
\centering
\caption{Spectroscopic information for \hii\, regions and candidates. } \label{tab:basic info}
\footnotesize   
\begin{tabular}{ccccccccccccc} 
\hline\hline
\centering
(1)             & (2)  & (3)     & (4)    & (5)      & (6)     & (7)     & (8)     & (9)           & (10)              & (11)            & (12)       & (13)  \\
WISE\_NAME      & Cat. & RA      & Dec    & Radius   & N2Ha    & S2Ha    & S2N2    & $V_{\rm LSR}$ & W$_{\rm H\alpha}$ & W$_{\rm [N II]}$ & Mod.       & Cls.  \\
                &      &         &        &          & (error) & (error) & (error) & (error)       & (error)           &  (error)        &            &       \\
                &      & [deg]   & [deg]  & [arcs]   &         &         &         & [km $\rm s^{-1}$] & [km $\rm s^{-1}$] & [km $\rm s^{-1}$] &      &       \\  
\hline
G085.325-00.789 & K    & 314.328 & 44.325 & 4308.3   & 0.264   & 0.127   & 0.483   & -0.493        & 25.315            & 18.292          & single     & HII  \\
                &      &         &        &          & (0.002) & (0.002) & (0.008) & (0.044)       & (0.109)           & (0.176)         &            &      \\
G086.814-00.141 & Q    & 315.014 & 45.875 & 402.7    & 0.342   & 0.268   & 0.782   & 2.850         & 27.571            & 22.3813          & single     & HII  \\
                &      &         &        &          & (0.005) & (0.008) & (0.027) & (0.081)       & (0.200)           & (0.443)         &            &      \\
G086.841+00.441 & C    & 314.405 & 46.277 & 772.0    & 0.396   & 0.209   & 0.527   & 5.048         & 21.760            & 14.294          & single     & HII  \\
                &      &         &        &          & (0.006) & (0.006) & (0.016) & (0.104)       & (0.254)           & (0.364)         &            &      \\
G087.223+00.347 & Q    & 314.869 & 46.504 & 63.5     & 0.413   & 0.224   & 0.543   & 7.103         & 22.954            & 14.471          & single     & HII  \\
                &      &         &        &          & (0.013) & (0.013) & (0.035) & (0.194)       & (0.474)           & (0.755)         &            &      \\
G088.059-00.021 & Q    & 316.078 & 46.887 & 75.5     & 0.386   & 0.325   & 0.843   & 9.554         & 28.260            & 22.807          & single     & HII  \\
                &      &         &        &          & (0.010) & (0.014) & (0.041) & (0.155)       & (0.383)           & (0.811)         &            &      \\
G088.340+01.796 & Q    & 314.327 & 48.296 & 190.1    & 0.454   & 0.290   & 0.639   & 7.880         & 24.352            & 31.976          & double     & HII  \\
                &      &         &        &          & (0.039) & (0.049) & (0.115) & (0.553)       & (1.359)           & (2.713)         &            &      \\
G088.647+01.827 & Q    & 314.594 & 48.549 & 120.3    & 0.371   & 0.691   & 1.862   & -1.847        & --               & --             & single     & HII  \\
                &      &         &        &          & (0.059) & (0.119) & (0.406) & (0.457)       & --           & --         &            &      \\
G088.674+00.313 & C    & 316.319 & 47.568 & 226.9    & 0.377   & 0.246   & 0.651   & 7.996         & 46.900            & 31.330          & single     & HII  \\
                &      &         &        &          & (0.010) & (0.010) & (0.029) & (0.284)       & (0.716)           & (0.923)         &            &      \\
\hline
\end{tabular}
\parbox{\linewidth}{
\justifying
\footnotesize 
\textbf{Notes.} Columns (1) $\sim$ (5) are provided by HIICat\_V2.3, among which Column (2) is the catalog suggested by \cite{Anderson2014ApJS} (``K" for Known, ``G” for Group, ``C” for Candidate and ``Q” for radio-Quiet). Columns (6) $\sim$ (11) are the spectroscopic information obtained in Section \ref{sec:spec process}. Column (12) is the Gaussian fitting method adopted for each source . Column (13) is the classification proposed by us in Section \ref{sec:identification}. Full Table is provided in the machine-readable format in the online Journal.
}
\end{table*}


\begin{table*}
\centering 
\caption{Physical parameters for \hii\ regions.} \label{tab:parameters}
\footnotesize
\begin{tabular}{ccccccc}
\hline\hline
\centering
(1)             &  (2)         & (3)         & (4)          & (5)     & (6)            & (7)       \\
WISE\_NAME      &  $T_{\rm e}$ & $n_{\rm e}$ & 12+log(O/H)  &  d      & $R_{\rm gal}$  & Flag\_$d$ \\
                &  (error)     & (error)     & (error)      & (error) & (error)        &           \\
                &  [K]         & [cm$^{-3}$] & [dex]        & [kpc]   & [kpc]          &           \\
\hline  
G085.325-00.789 &  7167.407    &  --         & 8.570        & 1.502   & 8.510          & maser/WISE \\
                &  (420.576)   &  --         & (0.002)      & (0.079) & (0.079)        &            \\
G086.814-00.141 &  6067.827    &  --         & 8.635        & 0.360   & 8.300          & kin/lamost \\
                &  (944.039)   &  --         & (0.004)      & (1.292) & (0.145)        &            \\
G086.841+00.441 &  6299.756    &  --         & 8.671        & 7.732   & 10.921         & OBstar     \\
                &  (791.596)   &  --         & (0.004)      & (2.717) & (2.717)        &            \\
G087.223+00.347 &  7429.771    &  --         & 8.681        & 0.313   & 8.244          & kin/lamost \\
                &  (1509.939)  &  --         & (0.008)      & (0.636) & (0.096)        &            \\
G088.059-00.021 &  6517.932    &  --         & 8.664        & 0.260   & 8.227          & kin/lamost \\
                &  (1943.710)  &  --         & (0.006)      & (0.217) & (0.047)        &            \\
G088.340+01.796 &  --          &  --         & 8.704        & 0.655   & 8.157          & OBstar     \\
                &  --          &  --         & (0.021)      & (0.046) & (0.046)        &            \\
G088.647+01.827 &  --          &  1292.658   & 8.655        & 2.174   & 8.368          & kin/lamost \\
                &  --          &  (1472.964) & (0.039)      & (1.455) & (0.172)        &            \\
G088.674+00.313 &  28505.833   &  --         & 8.658        & 0.188   & 8.317          & kin/lamost \\
                &  (3087.265)  &  --         & (0.007)      & (0.267) & (0.132)        &            \\                
\hline
\end{tabular}
\parbox{0.85\linewidth}{
\justifying
\footnotesize
\textbf{Notes.} Column (1) is the WISE\_Name ID. Columns (2) $\sim$ (5) are the physical parameters obtained in Section \ref{sec:results}. Column (7) indicates the method used to estimate distance, including ``maser/WISE", ``OBstar"', ``kin/lamost", and ``kin/wise". Full Table is provided in the machine-readable format in the online Journal. 
}
\end{table*}

\subsection{Distance and  Galactocentric distance} \label{sec:distance}

 In our sample, only 85 (33\%) \hii\ regions have distances provided by HIIcat\_V2.3. 
 Therefore, in this section, we determine distances for our sample. Two methods are employed: 1) the weighted-average distance of associated OB stars, and 2) the kinematic distance derived from $V_{\rm LSR}$. Afterwards, we supplement these distances with those from previous literature.
 
\subsubsection{Parallax distance}

 Spectrophotometric measurements of exciting stars have long served as a powerful tool for estimating distances to \hii\ regions \citep{Russeil2003A&A, Moises2011MNRAS, Foster2015AJ}. Thanks to advances in spectroscopic surveys, an increasing number of OB stars have been identified \citep{Liu2019ApJS}, which will help us to pinpoint the exciting sources of \hii\ regions. In recent years, several studies have estimated distances to \hii\ regions by cross-matching with OB stars \citep{Arellano_Cordova2020MNRAS, Mendez_Delgado2022MNRAS, Shen2025A&A, Zhang2025AJ}.

 Following a similar method, we combine the LAMOST OB star catalog \citep{Liu2019ApJS}, the Gaia OB stars catalog \citep{Xu2021A&A} and the OB stars retrieved from SIMBAD to obtain a merged OB star catalog. The OB stars from SIMBAD are selected and downloaded from the online site (\url{https://simbad.cds.unistra.fr/simbad/}), and selected based on spectral type. The merged OB catalog contains around 70,000 OB stars. We obtain distances for these OB stars from  \cite{Bailer_Jones2021AJ}, which provides parallax distances for 1.35 billion stars in Gaia Early Data Release 3 (EDR3).

 We cross-match the OB stars with \hii\ regions within their radius, and use the weighted average of the OB star distances as a proxy for the \hii\ region distances, where the weights are the angular distances from each OB star to the \hii\ region center. Outliers beyond 3 $\sigma$ from the median value are removed. 
 As a result, we determine OB distances ($d_{\rm OB}$) for 150 \hii\ regions, with an average uncertainty of 15\%. The other 105 \hii\ regions have not been cross-matched with OB stars, which is primarily because the current OB star catalog remains incomplete.

\subsubsection{Kinematic distance}

 The kinematic method utilizing $V_{\rm LSR}$ is also a common way to estimate distances to \hii\ regions. 
 We use the Monte Carlo calculation tool developed by \cite{Wenger2018ApJ} to estimate kinematic distances, adopting the new rotation curve and updated solar motion parameters provided by \cite{Reid2019ApJ}.
 
 Using the $V_{\rm LSR}$ obtained in Section \ref{sec:spec process}, we calculate kinematic distances for all 255 \hii\ regions.  We then exclude 50 sources within 15$^{\circ}$ of the Galactic center and 20$^{\circ}$ of the Galactic anti-center, as kinematic distances are unreliable in these directions due to velocity crowding \citep{Wenger2018ApJ}. Finally, we obtain reliable kinematic distances ($d_{\rm kin, lamost}$) for 205 \hii\ regions, with an average uncertainty of 13\%.

\subsubsection{Distance from archives}
 HIIcat\_V2.3 compiles distances for some sources from previous works. Most of these are kinematic distances, with a few being maser parallax distances. In our sample, 85 sources have distances provided by HIIcat\_V2.3 (hereafter $d_{\rm wise}$): 5 are maser parallax distances ($d_{\rm maser, wise}$), and 80 are kinematic distances ($d_{\rm kin, wise}$).
 
    \begin{figure}[htbp]
    \centering
    \includegraphics[width=\linewidth]{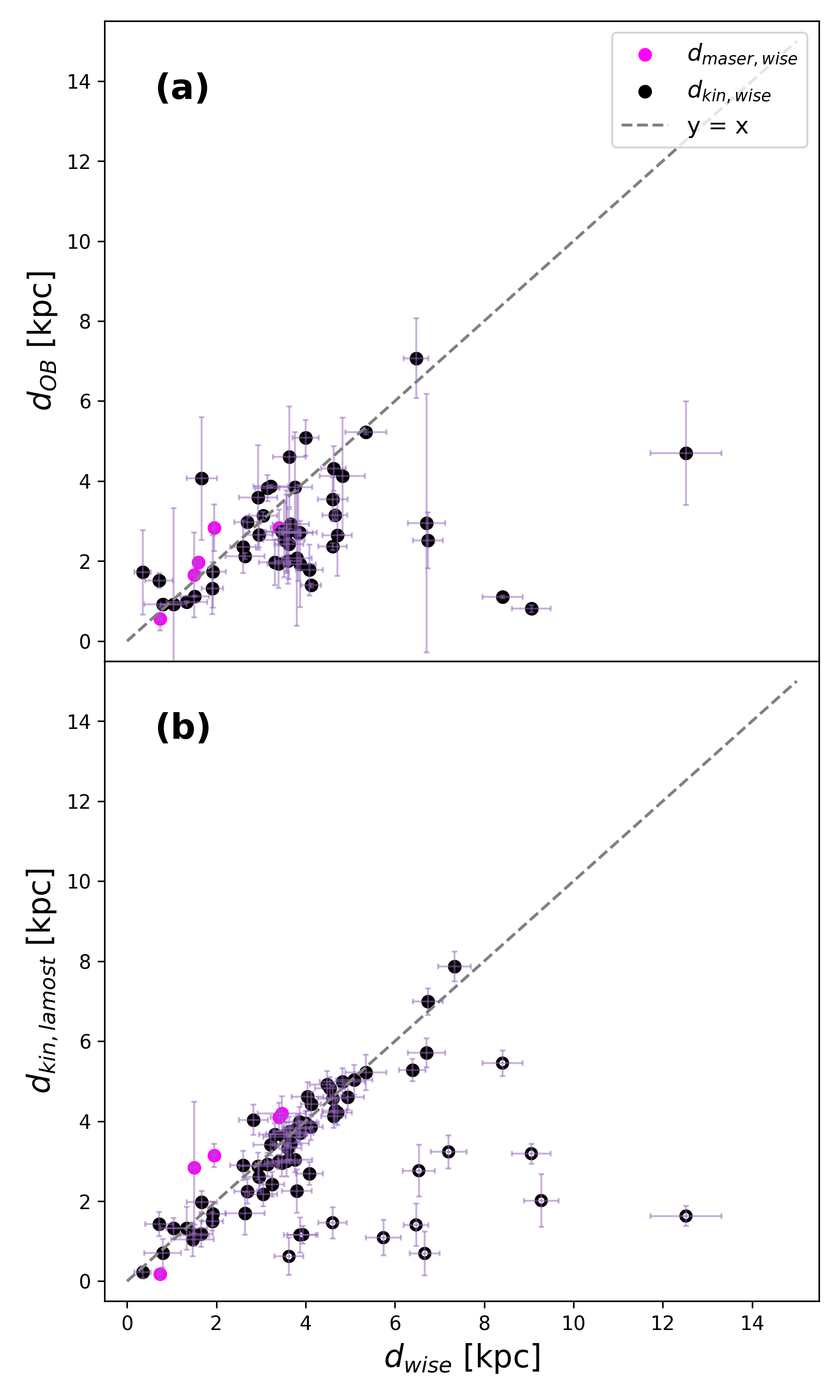}  
    \caption{Comparison of $d_{\rm OB}$, $d_{\rm kin, lamost}$ with $d_{\rm wise}$. The two panels share the same x-axis ($d_{\rm wise}$), with the y-axes corresponding to $d_{\rm OB}$ and $d_{\rm kin, lamost}$, respectively. Magenta dots denote maser parallax distances ($d_{\rm maser, wise}$), and black dots denote kinematic distances ($d_{\rm kin, wise}$) from HIIcat\_V2.3. Error bars are shown in purple. A gray dashed line of y = x is included in each panel for reference. Open circles in panel (b) indicate outliers relative to the y = x line.
    }
    \label{fig:dist_compare} 
    \end{figure}
    
 To validate our results, we compare $d_{\rm OB}$ and $d_{\rm kin, lamost}$ with $d_{\rm wise}$ in Figure \ref{fig:dist_compare}.
 Most of the distances are in good agreement, particularly the maser parallax distances (shown as magenta dots). 
 In panel (b), however, some outliers remain in the comparison of the two sets of kinematic distances, which are marked by open circles. 
 
\begin{figure}[htbp]
    \centering
    \includegraphics[width=0.95\linewidth]{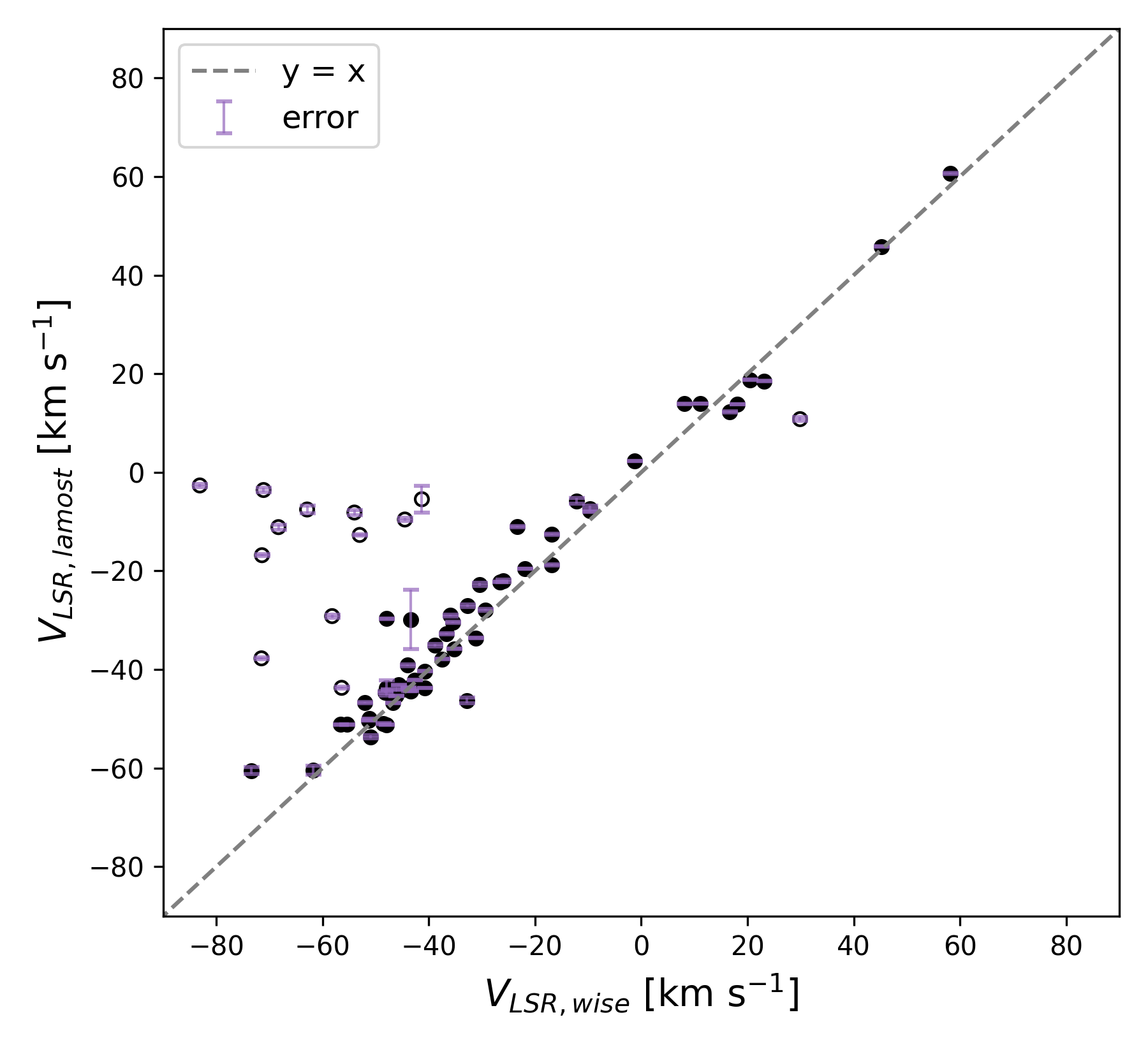}
    \caption{Comparison of $V_{\rm LSR, lamost}$ with $V_{\rm LSR, wise}$. $V_{\rm LSR, lamost}$ values are obtained from our LAMOST spectra in Section \ref{sec:spec process}. $V_{\rm LSR, wise}$ values are provided by HIIcat\_V2.3. Error bars are shown in purple. Open circles represent the outliers from panel (b) of Figure \ref{fig:dist_compare}. A gray dashed line y = x is included for reference.} 
    \label{fig:vlsr_compare} 
    \end{figure}

 We attribute these outliers to discrepancies in $V_{\rm LSR}$ values. To investigate this, we compare the $V_{\rm LSR, lamost}$ (obtained in Section \ref{sec:spec process}) with $V_{\rm LSR, wise}$ (from  HIIcat\_V2.3)  in Figure \ref{fig:vlsr_compare}. The outliers from panel (b) of Figure \ref{fig:dist_compare} are marked as open circles in Figure \ref{fig:vlsr_compare}. We find that these outliers also deviate from the 1:1 line in the $V_{\rm LSR}$ comparison.

\subsubsection{Adopted distance and Galactocentric distance}

 Ultimately, we combine the distances obtained from the above methods as our final distance ($d$). For \hii\ regions with multiple distances, we combine them according to the following priority order: $d_{\rm maser, wise} > d_{\rm OB} > d_{\rm kin, lamost} > d_{\rm kin, wise} $. Distance flags (`maser/WISE', `OBstar', `kin/lamost' and `kin/wise') are also provided to indicate which method was selected. In total, distances have been estimated for 243 \hii\ regions in our sample, including 6 $d_{\rm maser, wise}$, 145 $d_{\rm OB}$ and 92 $d_{\rm kin, lamost}$, respectively.  Additionally, we derive the Galactocentric distance ($R_{\rm gal}$) from $d$, adopting a Galactocentric distance of the sun of $\rm R_{\rm \odot} = 8.15\,kpc$ \citep{Reid2019ApJ}. Panel (i) of Figure \ref{fig:parahist} shows the histogram of $R_{\rm gal}$. Detailed values are provided in Table \ref{tab:parameters}.
 
\begin{figure*}[htbp]
    \centering
    \includegraphics[width=0.7\linewidth]{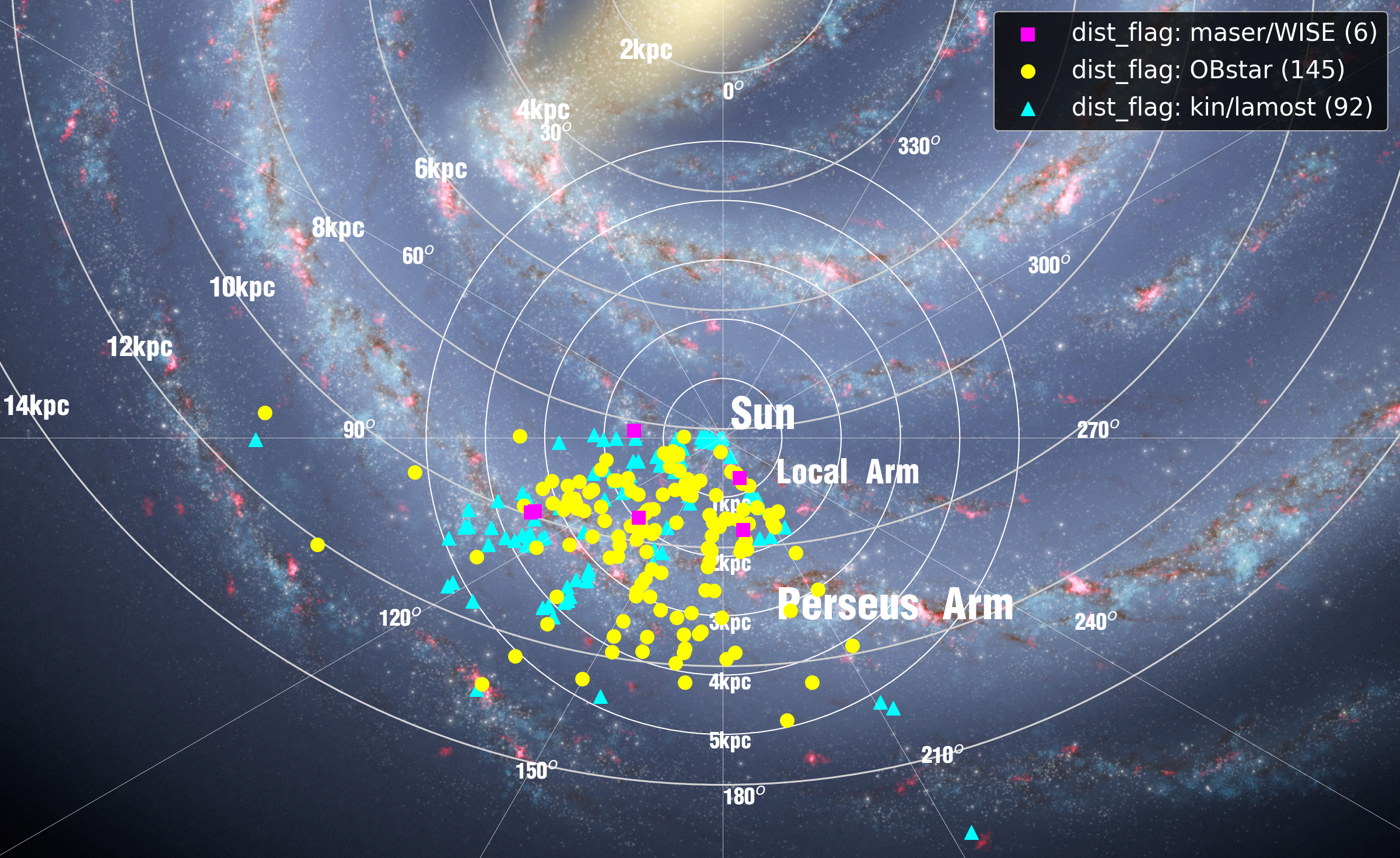}
    \caption{Face-on view of \hii\ regions in the Galactic plane. Galactic coordinates and the locations of the spiral arms are overlaid.
    Magenta squares, yellow dots, and cyan triangles indicate ``maser/WISE", ``OBstar", and ``kin/lamost", respectively.
    The background is an artistic impression of the Milky Way (Credit: NASA/JPL-Caltech/ESO/R.Hurt).}
    \label{fig:topview}
    \end{figure*}

 In Figure \ref{fig:topview}, we show a face-on view of the spatial distribution. The background image shows the spiral structure of the Milky Way. We indicate the different distance determination methods using distinct markers.  The \hii\, regions are mainly located in the Local Arm, the Perseus Arm, and the interarm regions between them.
 
\section{Results and Discussions} \label{sec:discussion}

\subsection{Radial Gradients of Line Ratios}

    \begin{figure*}[htbp]
    \centering
    \includegraphics[width=0.68\linewidth]{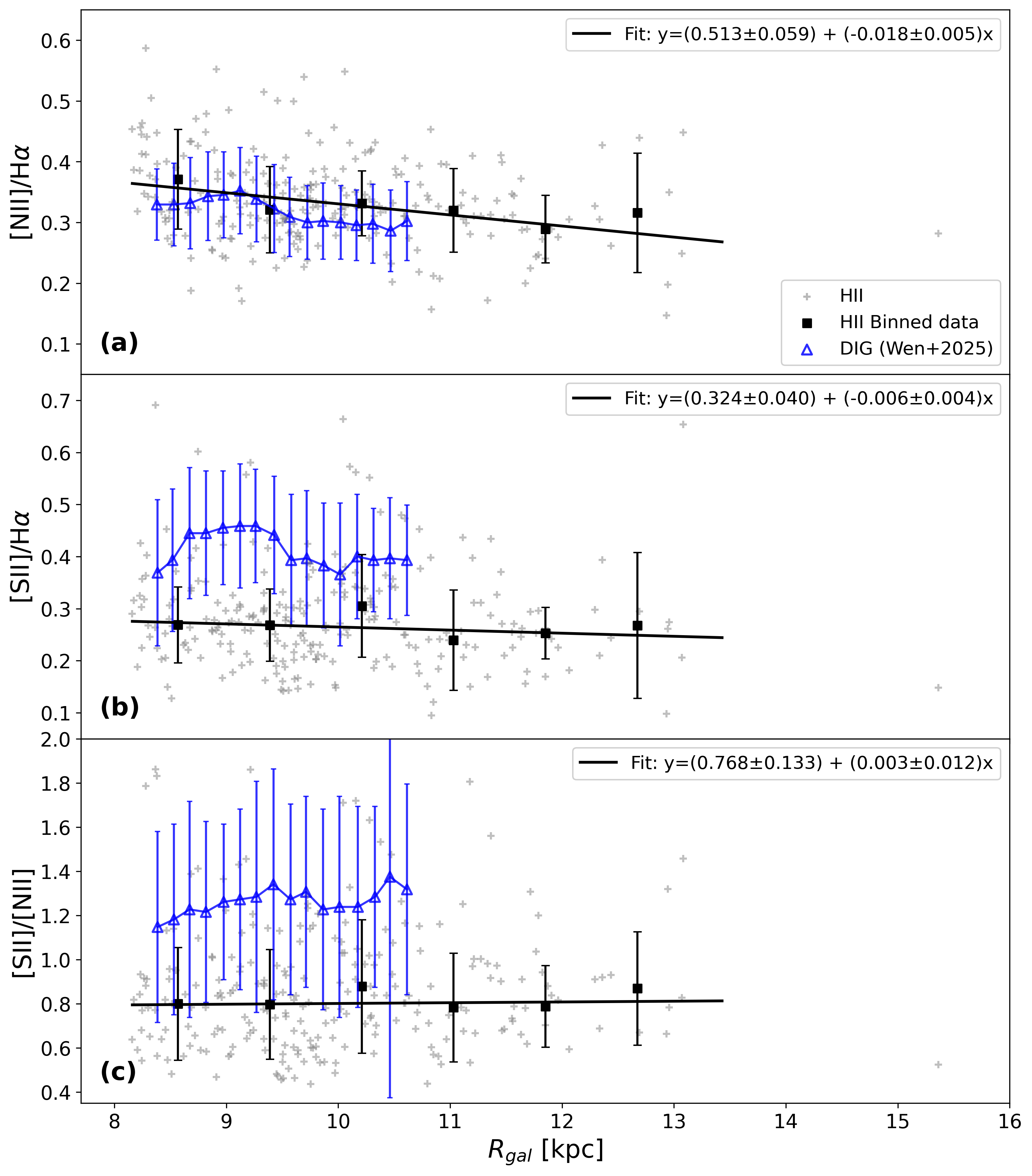}
    \caption{Radial gradients of line ratios as functions of $R_{\rm gal}$. From top to bottom, the panels show \nii/\ha, \sii/\ha, and \sii/\nii, respectively. Gray crosses denote \hii\ regions, and black squares represent the median values of binned \hii\ regions with scatter shown as error bars. Black lines correspond to linear fits to the median values, using errors as weights. Blue triangles and connecting  lines represent the DIG sample from \cite{Wen2025AJ}. }
    \vspace{\baselineskip}
    \label{fig:lineratio_gradiant} 
    \end{figure*}

 We plot the radial variation of the line ratios \nii/\ha, \sii/\ha\, and \sii/\nii\, as functions of $R_{\rm gal}$ in Figure \ref{fig:lineratio_gradiant}. 
 \hii\, regions are shown as gray crosses, and are binned according to $R_{\rm gal}$ shown in black squares. We perform a linear fit to the median values of each bin, and use the scatter of each bin as weights in the fitting. The best-fit linear relations are shown as black lines. We find that \nii/\ha\, and \sii/\ha\, show negative radial gradients with increasing $R_{\rm gal}$. The fitting slope of \sii/\nii\, is positive, but it is very shallow and the associated error is relatively large, making it a nearly flat radial distribution.

 We further compare our results with the DIG sample from \cite{Wen2025AJ}, plotted as blue triangles in Figure \ref{fig:lineratio_gradiant}.
 The radial distributions of line ratios in \hii\, regions are different from those in the DIG. The values of \nii/\ha\, are slightly higher than those in the DIG, while the values of \sii/\ha\, and \sii/\nii\, are significantly lower across all $R_{\rm gal}$. Differences in line ratios may reflect distinct excitation conditions. Compared with \hii\ regions, DIG is dominated by low-ionization states. In addition, variations in line ratios among different \hii\, regions also show possible correlations with the spectral types of their ionizing sources (Zhao et al., in prep.).
 
 In addition, \cite{Wen2025AJ} reported prominent peaks for \nii/\ha\, and \sii/\ha\, at $R_{\rm gal}$ $\approx$ 9.1 kpc, which they interpreted as a signature of distinct ionization mechanisms between spiral arm and inter-arm regions. In contrast, our \hii\, region sample does not show such peaks, indicating that there is no obvious difference between the \hii\, regions in the interarm region and those in the spiral arms. Furthermore, this result supports that the main ionizing source of the DIG in the interarm region might not be photons leaking from \hii\, regions. For \sii/\nii\, ratio, DIG exhibits a steeper slope in the inner disk and maintains near constant in the outer disk. \cite{Wen2025AJ} attributed this trend to the systematic variation in the ionization degrees (S$^+/$S) of DIG with $R_{\rm gal}$, and to a higher fraction of sulfur in the outer disk existing in the form of S$^{++}$. For \hii\, regions, \sii/\nii\, is nearly flat across $R_{\rm gal}$, implying that S$^+/$S varies only slightly with $R_{\rm gal}$.

\subsection{Radial Gradients of $T_{\rm e}$ and $n_{\rm e}$}

    \begin{figure*}[htbp]
    \centering
    \includegraphics[width=0.68\linewidth]{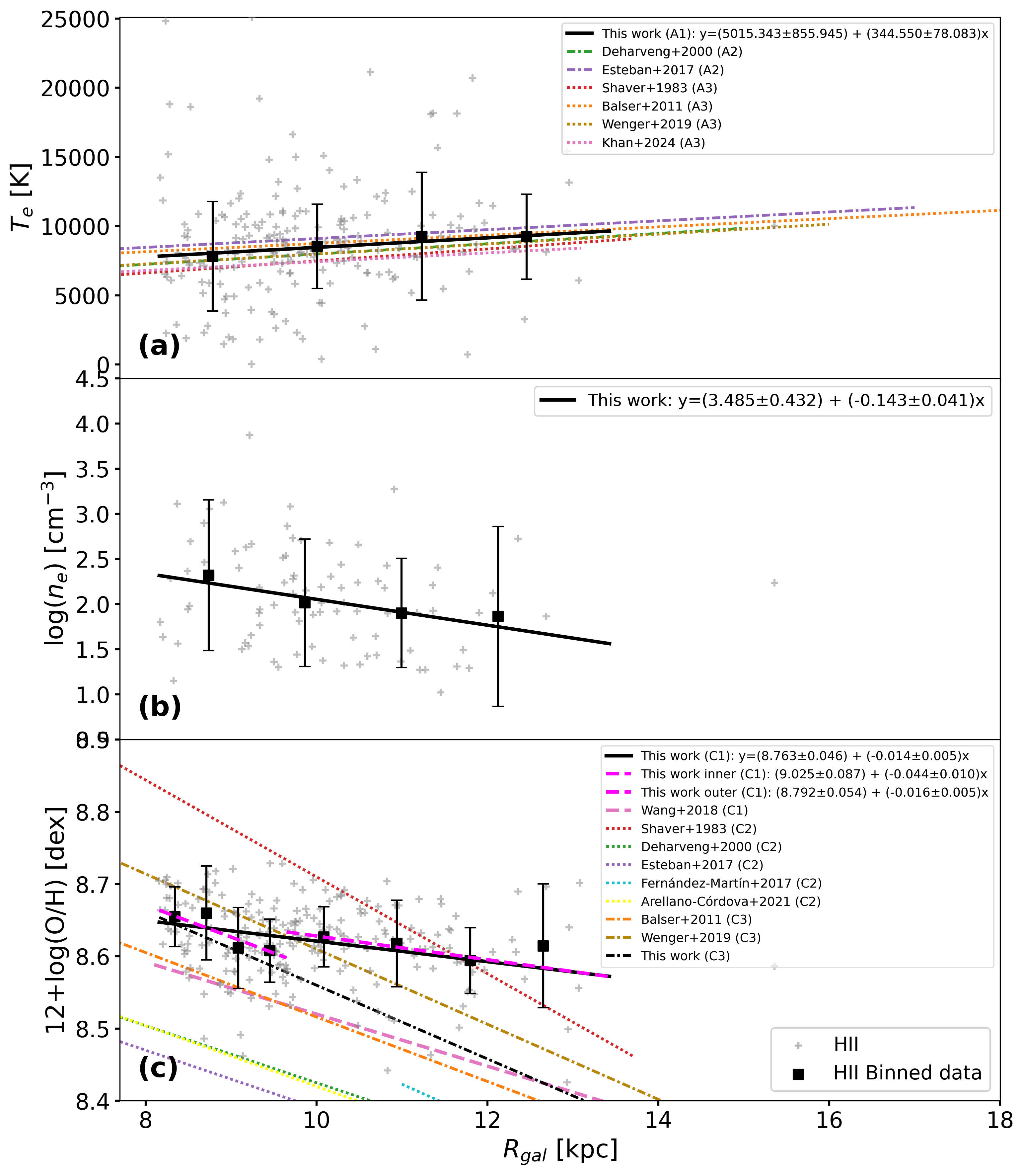}   
    \caption{Radial gradients of physical properties as functions of $R_{\rm gal}$. The panels from top to bottom are $T_{\rm e}$, $n_{\rm e}$ and oxygen abundance (in units of 12+log(O/H)), respectively. Gray crosses denote \hii\ regions, and black squares represent the median values of binned \hii\ regions with scatter shown as error bars. The black solid lines are the global fitting results for median values. 
    The other color lines are the gradients from the literature, with method labels: in panel (a) for $T_{\rm e}$, (A1) optical line width, (A2) optical auroral line ratio, and (A3) radio continuum flux and RRL; in panel (c) for 12 + log(O/H), (C1) N2\ha\, method, (C2) direct method, and (C3) empirical relationship between 12+log(O/H) and $T_{\rm e}$. In panel (c), the magenta dashed lines are the segment fitting results.}
    \label{fig:paramater_gradient} 
    \end{figure*}

 We present the $T_{\rm e}$ and $n_{\rm e}$ radial variations of \hii\, regions as functions of $R_{\rm gal}$ in panels (a) and (b) of Figure \ref{fig:paramater_gradient}.  The \hii\, regions are shown as gray crosses, and are binned according to $R_{\rm gal}$. 

 We perform linear fitting on the entire disk for $T_{\rm e}$, as shown by black solid line. 
 The $T_{\rm e}$ gradient is expressed as 
    \begin{equation}
        T_{\rm e} = 5015.343 (\pm855.945) + 344.550(\pm 78.083) R_{\rm gal}
    \end{equation}
 The slope (344.550 $\pm$ 78.083 K $\rm kpc^{-1}$) is consistent with previous works, such as \cite{Shaver1983MNRAS} (433 $\pm$ 40 K $\rm kpc^{-1}$), \cite{Deharveng2000MNRAS} (372 $\pm$ 38 K $\rm kpc^{-1}$), \cite{Balser2011ApJ} (299 $\pm$ 31 K $\rm kpc^{-1}$), \cite{Esteban2017MNRAS} (320 $\pm$ 40 K $\rm kpc^{-1}$), \cite{Wenger2019ApJ} (359 $\pm$ 20 K $\rm kpc^{-1}$) and \cite{Khan2024A&A} (372 $\pm$ 28 K $\rm kpc^{-1}$). 
 We compare our result with several previous works in panel (a) of Figure \ref{fig:paramater_gradient}, and label three common methods for measuring $T_{\rm e}$. We find there are no significant differences among the various methods.
 
 We also perform linear fitting on the entire disk for $n_{\rm e}$. 
 The $n_{\rm e}$ gradient is expressed as 
    \begin{equation}
        \log(n_{\rm e}) = 3.485 (\pm0.432) - 0.143 (\pm 0.041) R_{\rm gal}
    \end{equation}
 With increasing $R_{\rm gal}$, $n_{\rm e}$ is relatively lower. However, the dispersion is relatively large within the same $R_{\rm gal}$ range. This may indicate that beyond disk gradient, the local nebular conditions play an important role in shaping $n_{\rm e}$ \citep{Fernandez_Martin2017A&A, Esteban2017MNRAS}.

\subsection{Radial Gradient of Oxygen Abundance}

 We also plot the oxygen abundance radial variation of \hii\, regions as a function of $R_{\rm gal}$ in panel (c) of Figure \ref{fig:paramater_gradient}. As before, \hii\, regions are shown as gray crosses, and are binned according to $R_{\rm gal}$. We perform a weighted linear fit to the median values of \hii\ regions using the errors as weights, shown as black solid line in the figure. The radial oxygen abundance gradient is expressed as
    \begin{equation}
        12+\log(\mathrm{O/H}) = 8.763 (\pm0.046) -0.014(\pm 0.005) R_{\rm gal}
    \end{equation}
    
 In addition, we perform the segmented linear fits for the inner and outer disks separately, which a cutoff at $R_{\rm gal}$ = 9.65 kpc. The fit results are shown as magenta dashed lines, and expressed as 
 \newcounter{saveeqn}
 \setcounter{saveeqn}{\value{equation}}
 \stepcounter{saveeqn}
 \renewcommand{\theequation}{\arabic{saveeqn}.\arabic{equation}}
 \setcounter{equation}{0}
 \begin{align}    
     12+\log(\mathrm{O/H})
     = 9.025 (\pm & 0.087) - 0.044(\pm 0.010) R_{\rm gal} \nonumber     \\
                  & \text{for }R_{\rm gal} \leq 9.65 \text{kpc}          \\
     12+\log(\mathrm{O/H})
     = 8.792 (\pm & 0.054) - 0.016(\pm 0.005) R_{\rm gal} \nonumber     \\
                  & \text{for }R_{\rm gal} > 9.65 \text{kpc}
 \end{align}
 \setcounter{equation}{\value{saveeqn}}
 \renewcommand{\theequation}{\arabic{equation}}
 
 We compare our result with several previous works in panel (c) of Figure \ref{fig:paramater_gradient}, and label three common methods to determine oxygen abundance. The radial gradient in the inner disk (-0.044 $\pm$ 0.010 dex kpc$^{-1}$) agrees well with the determinations by \cite{Deharveng2000MNRAS} (-0.0395 $\pm$ 0.0049 dex kpc$^{-1}$), \cite{Balser2011ApJ} (-0.0446 $\pm$ 0.0046 dex kpc$^{-1}$) and \cite{Wang2018PASP} (-0.036 $\pm$ 0.004 dex kpc$^{-1}$). The radial gradient in the outer disk (-0.016$\pm$ 0.005 dex kpc$^{-1}$) is found to be flatter than that in the inner disk. The flattening trend has been proposed by some previous works, such as \cite{Fich1991ApJ, Vilchez1996MNRAS, Esteban2013MNRAS}. Several mechanisms may be responsible for the flattening: flattening of the star formation efficiency (SFE), radial metal mixing or enriched infall \citep{Bresolin2012ApJ}. \cite{Esteban2013MNRAS} have constructed a chemical evolution model and found the flattening of SFE can explain the chemical abundance flattening in the outer disk.
 
 Besides the radial gradient, the absolute values of oxygen abundance exhibit a large offset among these different works. Several factors may contribute to this discrepancy: (1) Differences in observational and data processing methods. We performed spectrum stacking within \hii\, regions, whereas most previous works only sampled the brightest central area using different fiber/aperture sizes. As \nii/\ha\, is not uniform within an \hii\, region but rather increases with distance from the center \citep{Zhang2025AJ}. Therefore, the oxygen abundance traced by \nii/\ha\, in the inner region is lower than that in the outer region of an \hii\, region, and thus the overall abundance obtained through stacking is higher than the central abundance. (2) Differences in \hii\, region samples and their positions within the Milky Way disk. There are indications that the radial gradient varies with azimuth \citep{Balser2011ApJ, Balser2015ApJ, Wenger2019ApJ}. (3) Differences in the methods used to estimate oxygen abundance. For example, some studies employ the N2\ha\, method \citep{Wang2018PASP}, some use the direct-$T_{\rm e}$ method \citep{Shaver1983MNRAS, Deharveng2000MNRAS, Esteban2017MNRAS, Fernandez_Martin2017A&A, Arellano_Cordova2021MNRAS}, while others calculate oxygen abundance from $T_{\rm e}$ using empirical relationships between 12 + log(O/H) and $T_{\rm e}$ \citep{Balser2011ApJ, Wenger2019ApJ}. 
 
 In the previous section, we check the $T_{\rm e}$ gradient and find it is similar to others. In order to examine the differences between methods, we transform our $T_{\rm e}$ gradient to 12+log(O/H) gradient using the empirical relationship by \cite{Shaver1983MNRAS}:
    \begin{equation}
        12+\log(\mathrm{O/H}) = 9.82 (\pm0.02) -1.49(\pm 0.11) \frac{T_{\rm e}}{10^{4}K}
    \end{equation}
 The transformed 12+log(O/H) gradient is as follows:
    \begin{equation}
        12+\log(\mathrm{O/H})' = 9.07 (\pm0.14) -0.051(\pm 0.012) R_{\rm gal}
    \end{equation}
 We also plot the transformed 12+log(O/H) gradient in panel (c) of Figure \ref{fig:paramater_gradient} as black dashed line to compare with other results. We find that the transformed gradient is similar to that in other works, and that the overall offset may be largely due to differences in methodology.

\subsection{2D Distributions of $T_{\rm e}$, $n_{\rm e}$ and Oxygen Abundance}

    \begin{figure*}[htbp]
    \centering
    \includegraphics[width=0.65\linewidth]{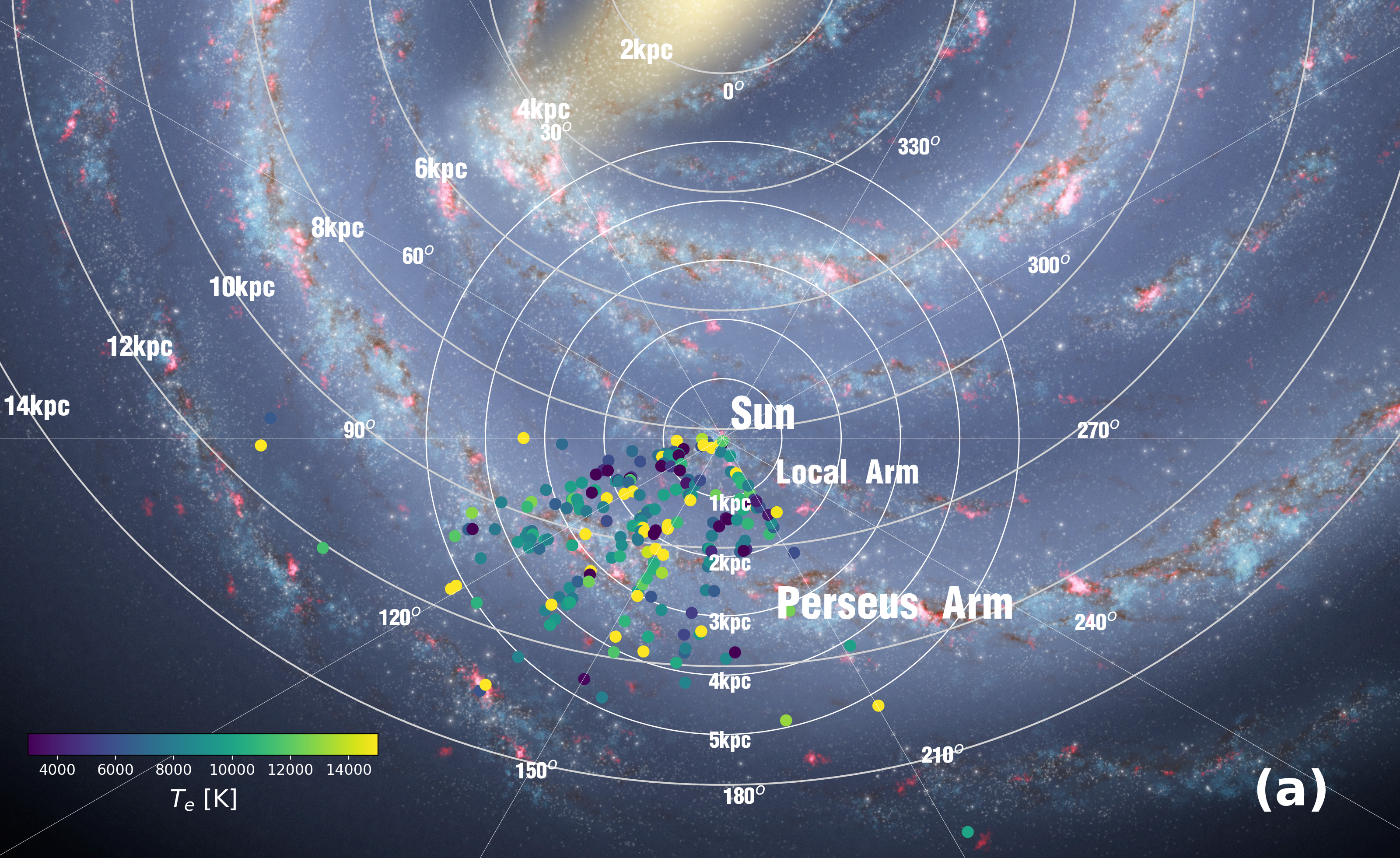}\\[6pt]
    \includegraphics[width=0.65\linewidth]{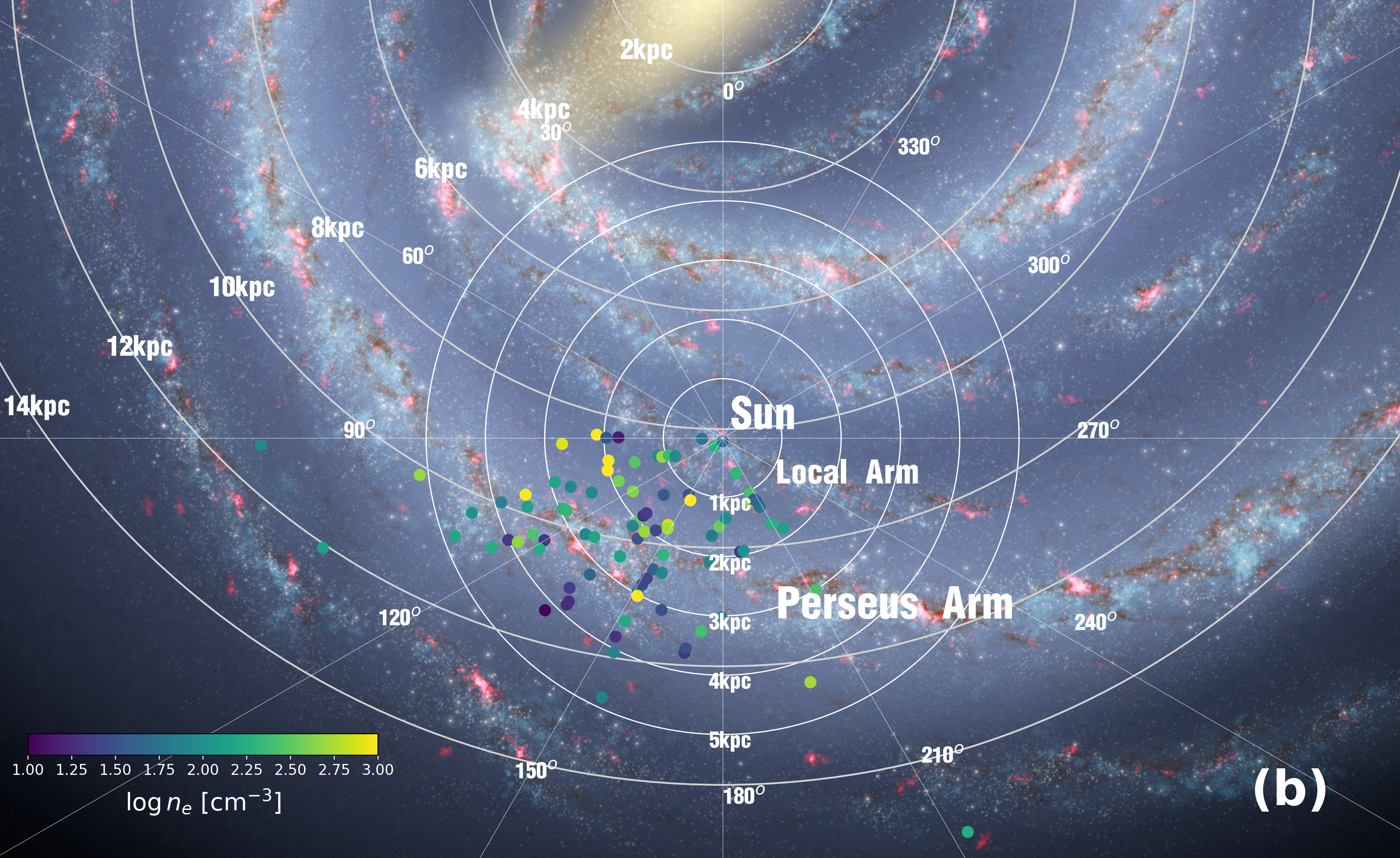}\\[6pt]
    \includegraphics[width=0.65\linewidth]{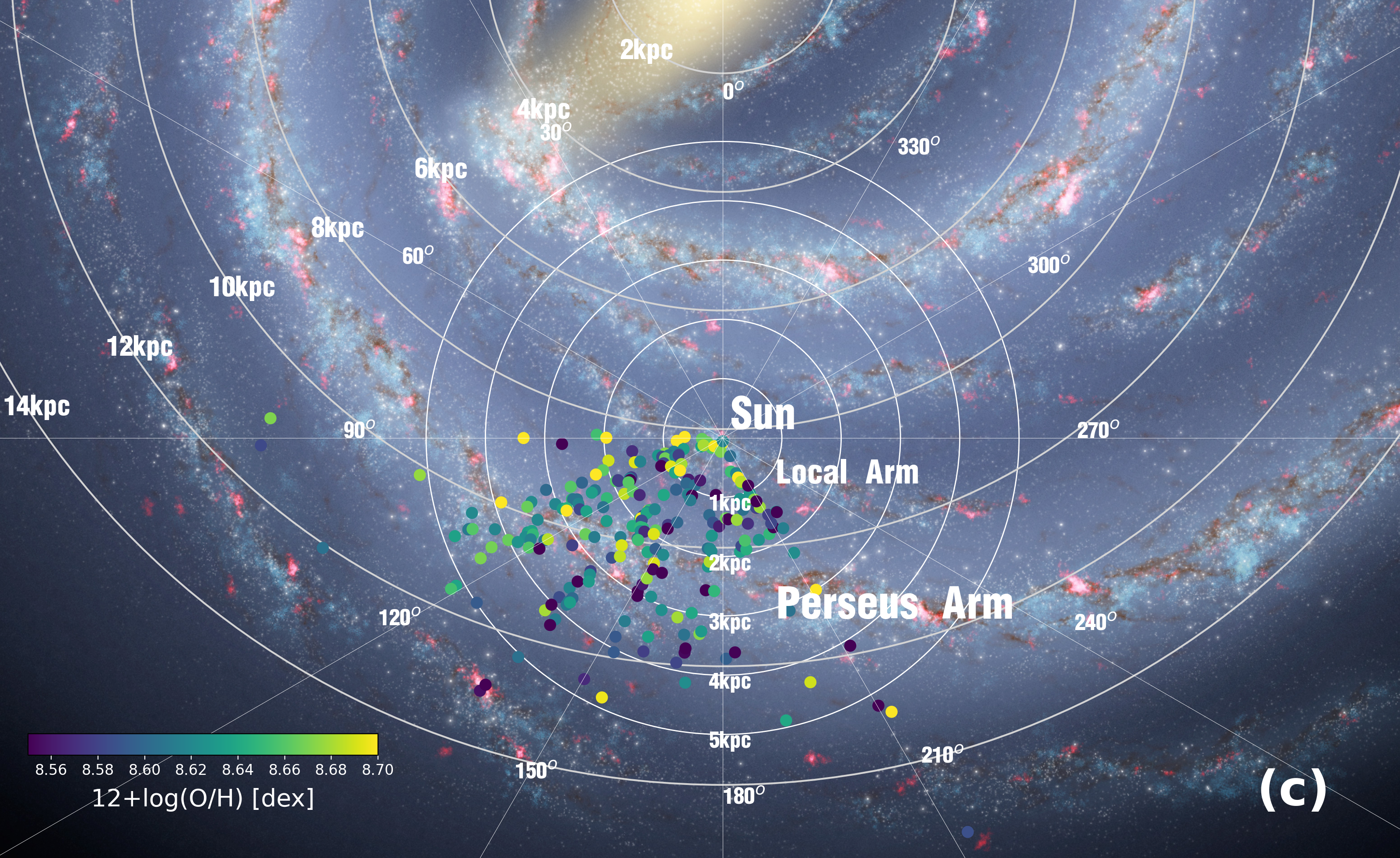}
    \vspace{1ex}
    \caption{2D distributions from face-on view of the Milky Way. The colors of points indicate the values of $T_{\rm e}$, $n_{\rm e}$, and $12+\log(\rm O/H)$. The color bars on the right show the range of each parameter. In three panels, the number of \hii\ regions is 227, 94, and 246, respectively. 
    }
    \label{fig:topview_contour}
    \end{figure*}

    \begin{figure}[htbp]
    \centering
    \includegraphics[width=\linewidth]{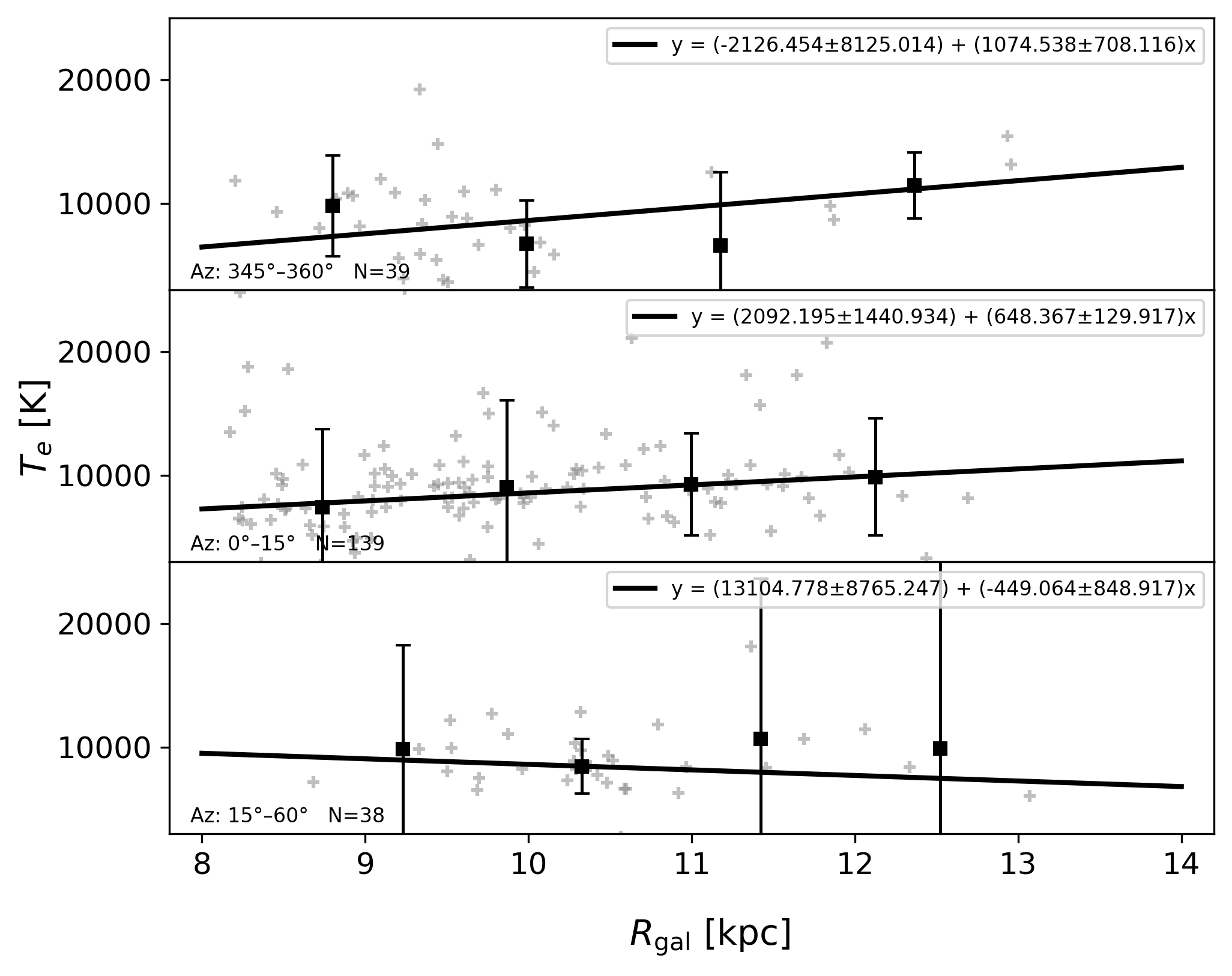}\\[6pt]
    \includegraphics[width=\linewidth]{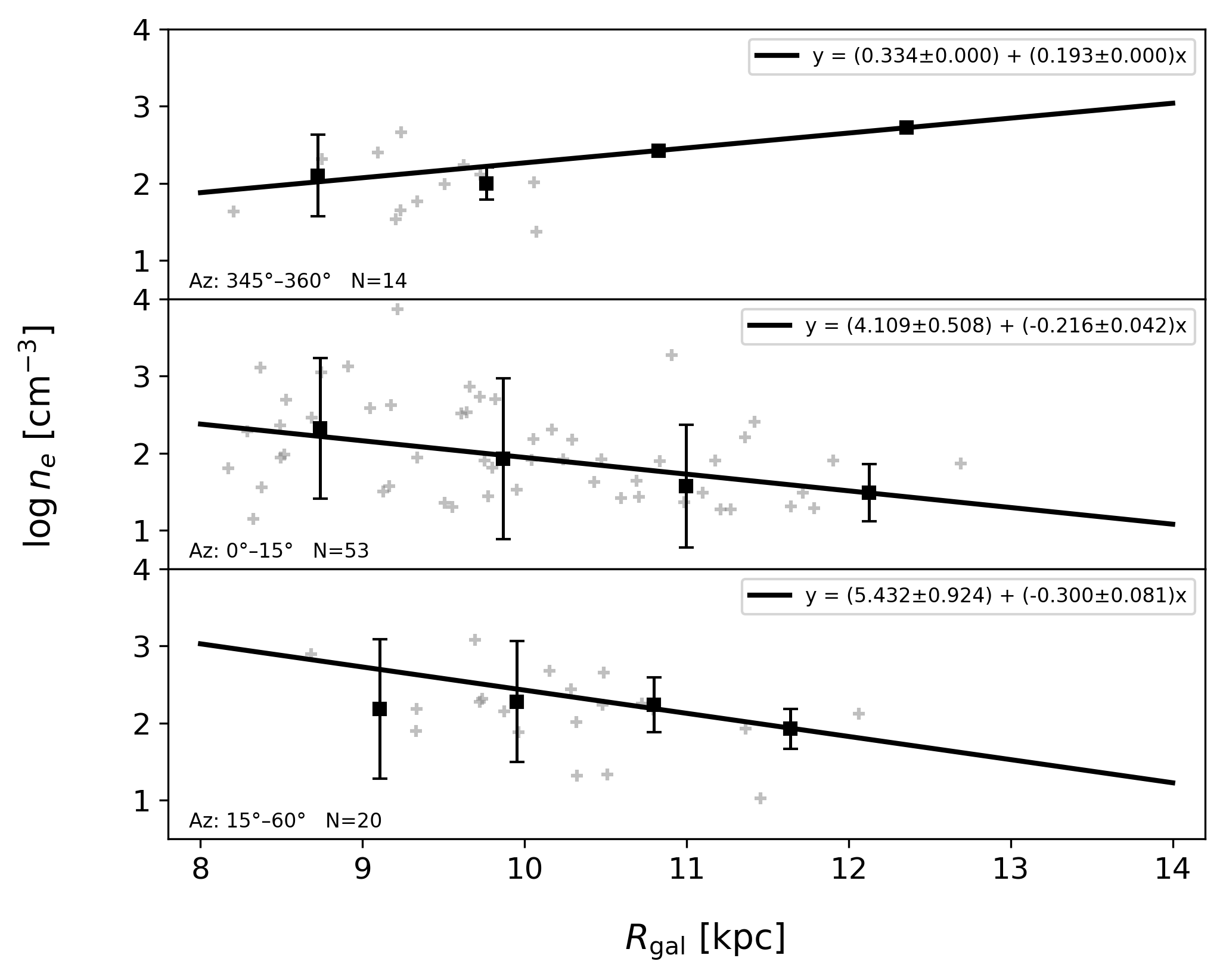}\\[6pt]
    \includegraphics[width=\linewidth]{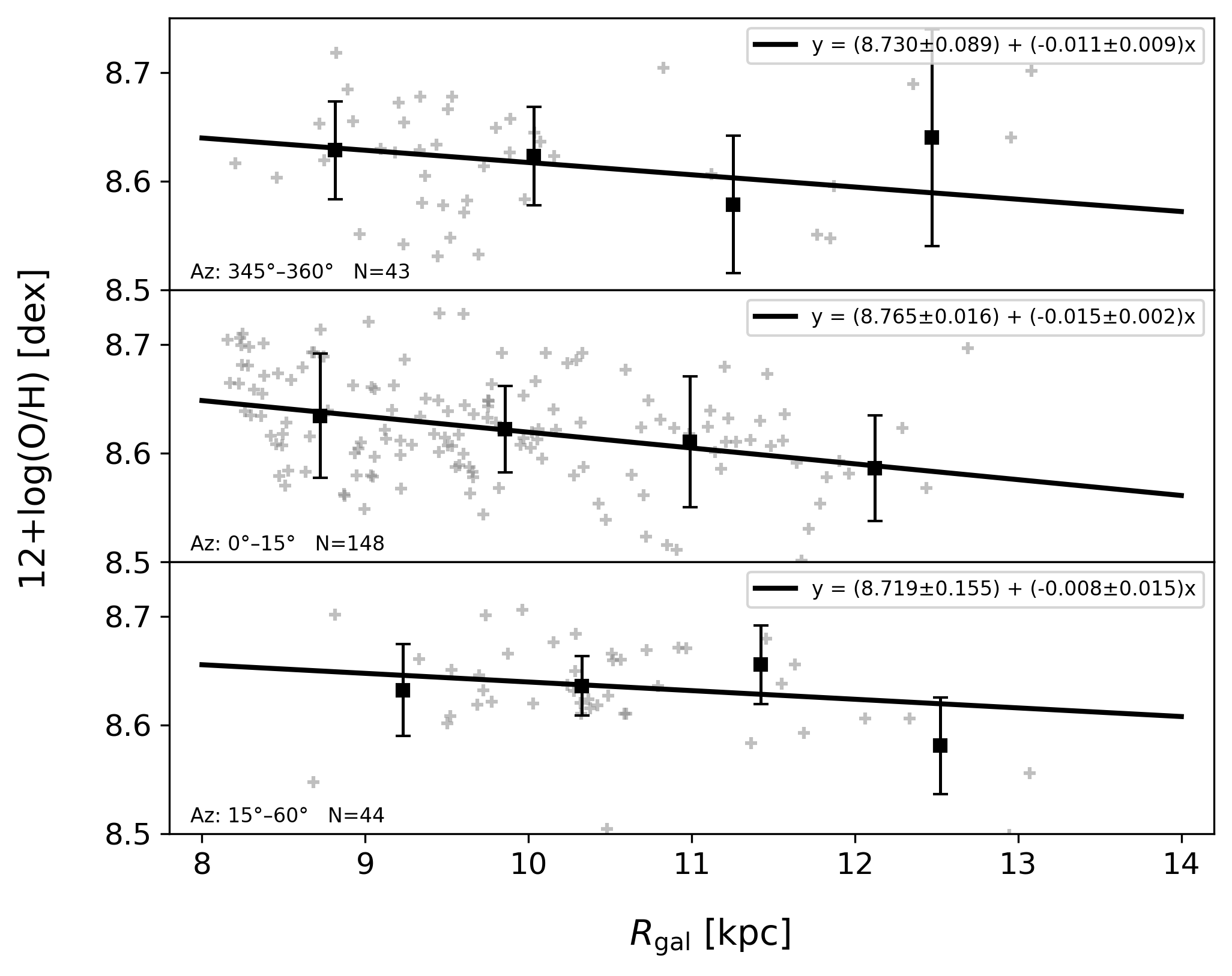}
    \vspace{1ex}
    \caption{Radial gradients of $T_{\rm e}$, $n_{\rm e}$, and $12 + \log(\rm O/H)$ at different azimuths. The panels from top to bottom are 345$^{\circ}$ $<$ Az $<$ 360$^{\circ}$, 0$^{\circ}$ $<$ Az $<$ 15$^{\circ}$, and 15$^{\circ}$ $<$ Az $<$ 60$^{\circ}$.
    Gray crosses denote \hii\ regions, and black squares represent the median values of binned \hii\ regions with errors representing the scatters in each bin. The black solid lines show the linear fitting results for the median values.}
    \label{fig:grandient_az}
    \end{figure}
    
 As the Galactocentric distance varies with Galactic longitude within a given spiral arm, \hii\, regions located on a spiral arm and those in the interarm region may have similar Galactocentric distances. Therefore, the radial distribution alone cannot reveal possible trends along a specific spiral arm or the potential differences between spiral arms and interarm regions. We thus present the two-dimensional distributions of $T_{\rm e}$, $n_{\rm e}$ and oxygen abundance on the face-on view of the Milky Way in Figure \ref{fig:topview_contour}. The \hii\, regions in this work are mainly located in the Local Arm, the Perseus Arm, and the interarm regions between them. We examine the \hii\, regions located in the arms, but no obvious trend is found along individual spiral arms in the 2D distributions. We note that non-circular motions can introduce additional uncertainties into kinematic distance estimates, especially in regions such as the Perseus Arm  \citep{Burton1971A&A, Gomez2006AJ, Reid2009ApJ, Reid2014ApJ, Reid2019ApJ}. These uncertainties may affect the assignment of some \hii\, regions to spiral-arm or inter-arm environments, and therefore may influence the inferred two-dimensional distributions of their physical properties. However, for sources with available OB-star parallax distances, we adopt these distances in preference to kinematic distances, which partly reduces this effect.

 We also examine the physical properties of \hii\, regions located in the arm and interarm regions, and find no significant systematic differences in  $T_{\rm e}$, $n_{\rm e}$ or 12+log(O/H). Instead, the spatial distributions of these quantities appear a patchy pattern and do not follow a simple arm–interarm distinction. Regarding this distinction, several studies have suggested that the physical properties of \hii\, regions remain consistent between arm and interarm environments \citep{Cedres2002A&A, Kreckel2016ApJ}. By contrast, other investigations based on larger galaxy samples have identified notable arm–interarm differences in certain galaxies \citep{Sanchez-Menguiano2017A&A, Sanchez-Menguiano2020MNRAS}. These studies further demonstrate that such differences are regulated by multiple galactic properties, including stellar mass, the presence of galactic bars, and spiral arm morphology (i.e., flocculent vs. grand-design spirals).

 In addition, we note that the radial gradients vary with azimuth. In Figure \ref{fig:grandient_az}, we divide the \hii\, regions into different sectors based on their azimuth and plot the radial gradients for each. In each sector, the \hii\ regions are binned and linear fitting is performed for the median values of each bin. The scatter with each bin is used as weights. The change in gradient slope across different azimuths can reach a factor of $\sim$ 2. In some sectors, the radial gradient even appears to be reversed; however, we consider this to be primarily due to the small number of data points in these sectors and large measurement errors. Azimuthal variations in $T_{\rm e}$ and $12+\log(\rm O/H)$ have been observed in both the Milky Way \citep{Balser2011ApJ, Balser2015ApJ, Wenger2019ApJ} and in other galaxies \citep[e.g.,][]{Sanchez-Menguiano2016ApJ, Sanchez-Menguiano2017A&A, Ho2017ApJ, Ho2018A&A}. Current explanations for such azimuthal variations involve streaming motions and radial migration induced by galactic bars \citep{Di_Matteo2013A&A}, by spiral arms \citep{Ho2017ApJ, Molla2019MNRAS, Spitoni2019A&A}, and by perturbation from minor galaxy interactions \citep{Bird2012MNRAS}.

\subsection{Distinguishing \hii\ Regions from DIG}

 DIG is an important component of the ionized gas in our Galaxy, which is estimated to account for approximately 90\% of it \citep{Reynolds1991IAUS, Haffner2009RvMP}. The ionizing sources for DIG are still under debate. Various possible ionizing sources include leaking photons from \hii\ regions \citep{Haffner2009RvMP}, hot low-mass evolved stars (HOLMES) \citep{Flores-Fajardo_2011, Zhang2017MNRAS}, and shocks \citep{Collins_2001}. Distinguishing DIG from \hii\ regions also remains a longstanding challenge, and a variety of methods have been widely adopted. These approaches primarily rely on the \ha\, surface brightness \citep{Zhang2017MNRAS, LiNiu_2021}, the \ha\, line equivalent width \citep{Belfiore_2016}, the \nii/\ha\, and \sii/\ha\, line ratios \citep{Kumari_2019}, or employing the classical BPT diagrams \citep{Baldwin1981PASP, Kauffmann_2003, Kewley_2006}.

\begin{figure}[htbp]
    \centering
    \includegraphics[width=\linewidth]{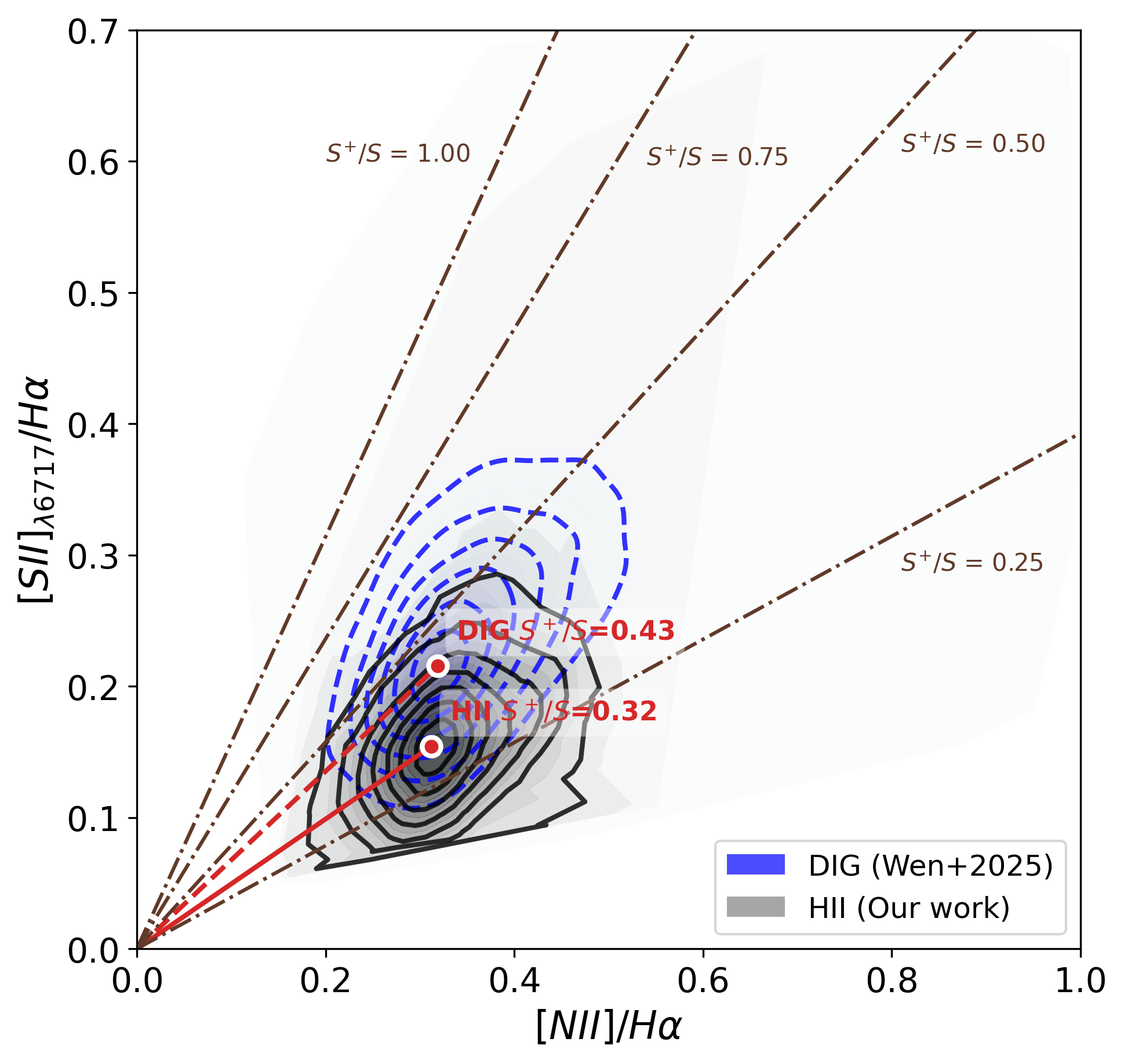}
    \caption{Distribution of HII regions and DIG on the \nii/\ha\, vs. \sii$\lambda$6717/\ha\, diagram. \hii\ regions are shown in gray contours. Blue contours represent DIG from \cite{Wen2025AJ}. The brown dash-dotted lines are the line emissivity models from \cite{Madsen2006ApJ}, corresponding to S$^+$/S = 0.25, 0.50, 0.75, and 1.00 from the lowest to the highest slope. The red solid and dashed lines indicate the density peaks of the \hii\, regions and DIG contour distributions, with corresponding S$^+/$S values of 0.32 and 0.43.} 
    \label{fig:ionization} 
    \end{figure}

 Benefiting from the high spatial resolution of LAMOST MRS-N, \citet{Wen2025AJ} selected DIG located outside \hii\ regions and SNRs. They constructed a sample of 17,821 DIG spectra in the anti-center region of the Milky Way and provided their line ratio information. Since both this DIG sample and the \hii\ regions in our work are selected from spatially resolved data, we then investigate their distributions in the \nii/\ha\, vs \sii$\lambda$6717/\ha\, diagram in Figure \ref{fig:ionization} to check whether these two populations can be distinguished. The \hii\, regions (gray contour) are concentrated at \nii/\ha\, = 0.31 and \sii$\lambda$6717/\ha\, = 0.15, while the DIG sample (blue contour) is concentrated at \nii/\ha\, = 0.32 and \sii$\lambda$6717/\ha\, = 0.22. Although the \hii\ regions have lower \nii/\ha\, and \sii$\lambda$6717/\ha\, ratios, they exhibit significant overlap with the DIG sample and cannot be effectively distinguished using the \nii/\ha\, vs. \sii$\lambda$6717/\ha\, diagram. This shows a discrepancy with Figure \ref{fig:lineratio_gradiant}, where \hii\, regions show slightly higher \nii/\ha, than the DIG in most radial bins. This difference is caused by the different comparisons used in the two figures: Figure 13 uses the full samples without radial binning, while Figure 9 compares the two samples within each $R_{\rm gal}$ bin; therefore, the full-sample distribution can be affected by the different $R_{\rm gal}$ coverage of the two samples. This also suggests that including radial information may make the classification between \hii\, regions and DIG more effective.
 
 We overplot the line emissivity models from \cite{Madsen2006ApJ} as dashed lines, representing ionization degrees (S$^+/$S) of 0.25, 0.50, 0.75, and 1.00, from the lowest to highest slope. The \hii\, regions show a shallower slope compared with the DIG sample and have a lower S$^+/$S ratio. We then estimate the specific S$^+/$S values for the \hii\, regions and the DIG sample, which are 0.32 and 0.43, respectively.

\section{Conclusion} \label{sec:conclusion}
 
 In this work, we combine LAMOST MRS-N data and the WISE \hii\, region catalog to construct a Galactic \hii\, region sample containing 280 \hii\, regions and candidates, 255 of which are further classified as \hii\, regions. We explore the radial gradients and 2D distributions of line ratios, $T_{\rm e}$, $n_{\rm e}$, and discuss the distinction between \hii\, regions and DIG based on line ratios, along with the limitations of using line ratios for such classification.. The main results are as follows:
 
 \begin{enumerate}    
    \item Based on stacked spectra of each \hii\, region and candidate, we measured four emission lines in our sample, including \ha, \nii$\lambda$6584, \sii$\lambda$6717, and \sii$\lambda$6731, and obtained spectroscopic information such as line ratios, FWHMs, and $V_{\rm LSR}$.

    \item Employing the diagnostic diagram of \nii/\ha\, and \sii/\ha, we identified 255 \hii\ regions in our sample, of which 90 are previously ``Known" \hii\ regions and 165 \hii\ region candidates are classified as \hii\ regions by us. The remaining 11 sources are considered PNe, and 14 fall within the SNR parameter ranges. For these 255 \hii\, regions, we derived physical parameters including $T_{\rm e}$, $n_{\rm e}$, oxygen abundances, and distances. 

    \item The radial distributions of \nii/\ha, \sii/\ha, and \sii/\nii\, in \hii\, regions differ from those in DIG. For \hii\, regions, \nii/\ha\, and \sii/\ha\, gradually decrease with increasing $R_{\rm gal}$, whereas for DIG, they show an upturn around $R_{\rm gal}$ $\sim$ 9.1 kpc. The gradient of \sii/\nii for \hii\, regions remains nearly flat in the entire disk, whereas DIG exhibits a steeper slope in the inner disk.

    \item The radial gradient of $T_{\rm e}$ can be described by a single linear fit with a slope of $344.550 \pm 78.083$ K kpc$^{-1}$. $n_{\rm e}$ exhibits a negative radial gradient with $R_{\rm gal}$, and the slope is $-0.143 \pm 0.041$ cm$^{-3}$ kpc$^{-1}$. The oxygen abundance shows different trends in the inner and outer disks. Specifically, it exhibits a steeper slope of $-0.044 \pm 0.010$ dex kpc$^{-1}$ in the inner disk and a shallower slope of $-0.016 \pm 0.005$ dex kpc$^{-1}$ in the outer disk. Furthermore, if we transform the $T_{\rm e}$ gradient into the oxygen gradient, it yields a single slope of $-0.051 \pm 0.012$ dex kpc$^{-1}$.
    
    \item 2D mapping reveals no clear systematic differences in the physical properties of \hii\, regions either between arm and interarm regions or along individual spiral arms. However, significant azimuthal variations in the radial gradients are observed, with slope changes reaching up to a factor of $\sim$2 across different sectors.

    \item H II regions have a median S$^+$/S value of 0.32, which is lower than that of the DIG (S$^+$/S = 0.43). Furthermore, owing to significant overlap in the \nii/\ha\, vs \sii$\lambda$6717/\ha\, diagram, \hii\, regions and the DIG cannot be effectively distinguished using this diagram alone.
\end{enumerate}

In addition to the above analyses, the constructed sample of 255 spectroscopically confirmed Galactic \hii\ regions also provides a valuable foundation for future statistical investigations. With the optical emission line measurements, we plan to explore internal turbulence and stellar feedback within \hii\ regions, as well as their evolutionary trends. Specifically, we aim to examine the size–velocity dispersion, the luminosity–velocity dispersion relations, and their dependence on Galactocentric distance and local environments. These studies will further our understanding of the dynamical states and physical processes governing \hii\ regions across the Milky Way disk.

\begin{acknowledgments}
We thank the anonymous reviewer for constructive comments and suggestions, which helped improve the clarity and quality of this manuscript. This work is supported by the National Key R\&D Program of China grant (Nos. 2021YFA1600401 and 2021YFA1600400) and the National Natural Science Foundation of China (NSFC, Nos. 12090041, 12090044, and 12090040). ZZ is supported by the National Key R\&D Program of China grant (Nos. 2023YFC2206403, 2024YFA1611602) and NSFC grant (Nos. 12373012 and 12041302),  as well as CMS-CSST-2025-A08. AYY acknowledges the support from the National Key R$\&$D Program of China grant (No. 2023YFC2206403), National SKA Program of China (No. 2025SKA0140100), and National Natural Science Foundation of China (No. 12303031). JJR thanks the support of the China Manned Space Program (No. CMS-CSST-2025-A19). This work is also sponsored by the Strategic Priority Research Program of the Chinese Academy of Sciences (No. XDB0550100). This work made use of the data from LAMOST (Large Sky Area Multi-Object Fiber Spectroscopic Telescope, also known as the Guoshoujing Telescope) (https://cstr.cn/31118.02.LAMOST). LAMOST is a Chinese national mega-science facility, operated by National Astronomical Observatories, Chinese Academy of Sciences.

\end{acknowledgments}

\bibliography{ref}{}
\bibliographystyle{aasjournal}

\end{document}